\documentclass[ prx, twocolumn, nofootinbib, floatfix]{revtex4-2}

\usepackage{notes2bib}
\usepackage{natbib}
\usepackage{gensymb}
\usepackage{amsthm}
\usepackage{amsmath}
\usepackage{color}
\usepackage{bm}
\usepackage{amsmath,amssymb}
\usepackage{amsfonts}
\usepackage{dsfont}
\usepackage{graphicx}
\usepackage{float}
\usepackage{subfigure}
\usepackage{verbatim}
\usepackage{amsfonts}
\usepackage{bbold}
\usepackage{dcolumn}
\usepackage{bm}
\usepackage[dvipsnames]{xcolor}
\usepackage{listings}
\definecolor{blueprl}{RGB}{46,48,146}
\usepackage[pdftex,colorlinks=true,urlcolor=blueprl,citecolor=blueprl,linkcolor=blueprl]{hyperref}
\usepackage{times}
\usepackage{color}
\usepackage[dvipsnames]{xcolor}

\usepackage[mathscr]{euscript}
\usepackage{amsmath}
\usepackage{graphicx}
\usepackage{dcolumn}
\usepackage{bm}
\usepackage{epsfig}
\usepackage{amssymb,latexsym,mathrsfs}
\usepackage{graphicx}
\usepackage{color}
\usepackage{hyperref}
\usepackage{float}
\usepackage{diagbox}

\newcommand{\vt}{\vphantom{\frac{1}{2}}}

\definecolor{vividviolet}{rgb}{0.62, 0.0, 1.0}
\definecolor{amaranth}{rgb}{0.9, 0.17, 0.31}
\definecolor{palatinateblue}{rgb}{0.15, 0.23, 0.89}
\definecolor{brightpink}{rgb}{1.0, 0.0, 0.5}
\definecolor{cornflowerblue}{rgb}{0.39, 0.58, 0.93}
\definecolor{deepcarminepink}{rgb}{0.94, 0.19, 0.22}
\definecolor{radicalred}{rgb}{1.0, 0.21, 0.37}
\definecolor{blueblue}{RGB}{21,47,181}
\definecolor{greengreen}{RGB}{65,166,16}

\newcommand{\be}{\begin{equation}}
\newcommand{\ee}{\end{equation}}
\newcommand{\bs}{\begin{split}} 
\newcommand{\bea}{\begin{eqnarray}}
\newcommand{\eea}{\end{eqnarray}}
\newcommand{\non}{\nonumber }

\usepackage{mathtools}

\newsavebox{\myhbar}
\newcommand{\dd}{\dagger}
\newcommand{\VT}{\vphantom{\frac{1}{\sqrt{2}}}}

\begin{document}

%\title{(1+1)-dimensional Hawking radiation model for a Kerr-Newman black hole} 
\title{Telefilters, telemirrors, and causality}

\author{Joshua Foo}
\email{joshua.foo@uqconnect.edu.au}
\affiliation{Centre for Quantum Computation \& Communication Technology, School of Mathematics \& Physics, The University of Queensland, St.~Lucia, Queensland, 4072, Australia}
\author{Sho Onoe}
\affiliation{Centre for Quantum Computation \& Communication Technology, School of Mathematics \& Physics, The University of Queensland, St.~Lucia, Queensland, 4072, Australia}
\author{Magdalena Zych}
\affiliation{Centre for Engineered Quantum Systems, School of Mathematics and Physics, The University of Queensland, St. Lucia, Queensland, 4072, Australia}
\author{Timothy C. Ralph}
\email{ralph@physics.edu.au}
\affiliation{Centre for Quantum Computation \& Communication Technology, School of Mathematics \& Physics, The University of Queensland, St.~Lucia, Queensland, 4072, Australia}

\begin{abstract}
We present new theoretical models for quantum optical mode-selective filters and mirrors using continuous-variable teleportation. We call these devices telefilters and telemirrors respectively. Both devices act as the identity channel on a mode of interest from an input multi-mode field while filtering or reflecting all the orthogonal modes. We utilise these models to analyse a causality problem in relativistic quantum optics, specifically the apparently acausal transmission and propagation of temporally delocalised wavepackets through mode-selective mirrors. Firstly, we show how telemirrors -- and thus mode-selective operations generally -- enact a fundamental time-delay on such wavepackets, which is necessary in order to prevent violations of causality. In an attempt to circumvent this delay, we next consider teleporting the independent temporal components of the input field separately and continuously, that is, performing operations `on-the-run'. In this scenario, the telemirror transmits the mode of interest as well as orthogonal modes which carry with them uncorrelated noise. In this scenario, the device may be considered mode-discriminating but not mode-selective. 
\end{abstract} 

\date{\today} 

\maketitle

\section{Introduction}
Unitary operations in quantum optics are defined by their action on single modes, or between multiple modes of the electromagnetic field \cite{bachor2004guide,walls2007quantum}. However, physical devices are usually only weakly selective of the modes which they act upon. For example, a beamsplitter will typically allow a large number of different mode pairs to be mixed.

Mode-discriminating interactions \cite{Rohde_2007} are more difficult to arrange physically but play an important role in quantum optics and interacting quantum field theories. Taking as input a multi-mode field, a mode-discriminating unitary affects a specific mode (for example, a spectral pulse) differently compared to modes orthogonal to it. A special class of mode-discriminating interactions is a mirror which transmits a single mode from a multi-mode input field and filters out or reflects all others. We refer to this specific kind of interaction where only one mode is \textit{transmitted} as mode-selective, to be distinguished from a mode-discriminating interaction which allows all modes to be transmitted while uniquely affecting one mode (conversely, affecting all but one mode). Mode-selective mirrors and beamsplitters have been identified as key tools in future communications and metrology applications \cite{ ansariPRXQuantum.2.010301,DonohuePhysRevLett.121.090501,WasilewskiPhysRevLett.99.123601,polinodoi:10.1116/5.0007577,macleanPhysRevA.100.033834}. Other mode-discriminating interactions such as phase-shifters, displacements and squeezers have been studied in the context of relativistic quantum communication protocols \cite{suPhysRevD.90.084022,onoePhysRevD.98.036011,onoePhysRevD.99.116001,FooPhysRevD.101.085006,Foo_2020, SuPhysRevX.9.011007} describing interactions between quantum fields and observers in relativistic reference frames. 

Mode-selective operations have been realized experimentally. One notable example is the quantum pulse gate \cite{Eckstein_2011} which overlaps a weak input field with a strong classical gating pulse in a nonlinear crystal. Using a technique known as spectrally engineered sum-frequency generation, only the mode from the input field which matches the form of the gating pulse is converted into a wavepacket at their sum frequency \cite{ansariPhysRevA.96.063817,Christ_2013,Brecht_2011,ansariPhysRevLett.120.213601}. A frequency filter can then select the mode of interest. Another example is the Raman quantum memory \cite{reimPhysRevLett.108.263602}, which interacts a train of control pulses with a signal mode inside an atomic vapour cell, yielding a controllable output which can be implemented as an optical beamsplitter network \cite{Reddy:17,bussieresdoi:10.1080/09500340.2013.856482,guoPhysRevA.97.063805}.

In this paper, we propose a new protocol for implementing the quantum optical mode-selective mirror, using continuous-variable (CV) teleportation \cite{BennettPhysRevLett.70.1895,vaidmanPhysRevA.49.1473,BraunsteinPhysRevLett.80.869}. We first consider a general model which demonstrates a {\it telefilter} which transmits a specific single mode whilst blocking all others, and then a {\it telemirror} which transmits a specific single mode whilst reflecting all others. 

Next, we develop a simplified model for a temporally delocalised wavepacket, where an input state is distributed over two independent temporal modes \cite{BrechtPhysRevX.5.041017,Ansari:18,Raymer_2020} which interact with the mirror at earlier and later times (this is then extended to $N$ temporal modes in Appendix \ref{sec:nmode}). Our motivation in studying this particular case is to understand how the interaction between temporally delocalised modes within mode-selective mirrors preserves causality, since such causal considerations are commonly neglected in theoretical models \cite{Christ_2013}. Our results show that a fundamental time-delay on the propagation of input modes is necessary in order for the action of such mirrors (and mode-selective mirrors generally) to remain consistent with relativity and to avoid violations of causality. Essentially, both the telefilter and telemirror enact a measurement of the input wavepacket, the duration of which is constrained by the temporal length of the wavepacket itself. Obviously, performing such a measurement instantaneously is unphysical. Our results thus highlight some tacit assumptions within quantum optical models for mode-selective interactions such as the Schmidt decomposition \cite{Rohde_2007}, which typically neglect time-ordering effects \cite{quesadaPhysRevA.90.063840,Christ_2013} and hence, admit acausal solutions. In line with this, our approach may be considered a quantum optical perspective on the important and ongoing discussion of causality-violating scenarios in relativistic quantum field theory, for example in the context of Sorkin's `impossible' measurements \cite{AharonovPhysRevD.34.1805, PopsecuPhysRevA.49.4331,Sorkin:1993gg,BeckmanPhysRevD.65.065022,Lin:2013loa,fewster2020quantum,ramonPhysRevD.103.085002,martinPhysRevD.103.025007}.

Finally, we propose an alternative teleportation model for the mirror \textit{without} a fundamental time-delay, wherein the individual temporal components are teleported separately and continuously before being coherently recombined at an output port. We show that in general this model fails as a mode-selective mirror, instead allowing the selected mode \textit{and} modes with the same temporal support but orthogonal to it to be transmitted. However, these orthogonal modes also carry additional sources of uncorrelated noise. When this noise is sufficiently large (i.e.\ the quadrature amplitudes are much larger than the quantum uncertainty limit), a signal encoded within these modes cannot be extracted by an observer on the receiving side of the mirror. In this version of the mirror with no delay, a type of mode-discrimination is achieved. These properties of the no-delay telefilter and telemirror may have important consequences in applications such as quantum communication and quantum causality \cite{Oreshkov:2011er}. 

Our paper is organised as follows: in Sec.\ \ref{sec:II}, we review the basic theoretical construction of mode-selective mirrors and the desired effect that they have upon an input multi-mode field. In Sec.\ \ref{sec:CVII}, we review two approaches to continuous-variable (CV) teleportation. We then show how these teleportation protocols can function as mode-selective devices wherein a single mode of interest from an input field is transmitted to a receiver. In Sec.\ \ref{sec4causality}, we highlight the neglect of causality considerations in models for mode-selective mirrors, which motivates us to generalise the telefilter and telemirror models to an input of two independent temporal modes. We introduce and study this new model in Sec.\ \ref{timedelay}, showing that it enacts an unavoidable time-delay upon the mode of interest as it propagates between the sender and receiver. In Sec.\ \ref{causal}, we propose another model for the temporal mode telemirror and telefilter where the temporal components are teleported separately and continuously. We offer some conclusions and implications of our results in Sec.\ \ref{sec:VI}. 

Throughout this paper, we utilise natural units, $\hslash = c = 1$.

\section{Mode-selective mirrors}\label{sec:II}
Before introducing our new model for the mode-selective mirror, we review a standard theoretical implementation of such a mirror and the transformation it enacts upon incoming field modes. 

Firstly, consider a complete set of orthonormal modes, $\{\hat{a}_l\}$, within which the mode of interest, $\hat{a}_0$, is contained. Likewise, we introduce a matching set of orthonormal modes, $\{\hat{b}_l\}$ which are orthogonal to $\{\hat{a}_l\}$ and contain the mode $\hat{b}_0$, which is complementary to $\hat{a}_0$. By matching, we mean that $\{\hat{a}_l\}$ and $\{\hat{b}_l\}$ are in the same spatiotemporal mode, enabling them to interact at a beamsplitter. We wish to construct a mode-selective unitary which only affects the mode of interest, $\hat{a}_0$, and its complement incident from the other direction, $\hat{b}_0$. 

To highlight the effect of a mode-selective unitary on the incident modes $\{\hat{a}_l\}$, $\{\hat{b}_l\}$, it is instructive to contrast this with that of a passive, or non-mode-selective unitary. A non-mode-selective interaction can be modelled via the unitary, 
\begin{align}\label{unitary1}
    \hat{U} &= \exp \bigg( - i \theta \bigg( \sum_l \hat{a}_l\hat{b}_l^\dd + \text{H.c} \bigg) \bigg) .
\end{align}
Consider an incoming mode $\hat{a}_\text{in}$, constructed as a superposition of the modes in $\{\hat{a}_l\}$,
\begin{align}
    \hat{a}_\text{in} &= \sum_l f_l \hat{a}_l
\end{align}
where $\sum_l |f_l|^2 = 1$, interacting with the other set of incident modes, $\{\hat{b}_l\}$. If we prepare the single-photon state $\hat{a}_\text{in}^\dd | 0 \rangle$ (where $| 0 \rangle$ is the state annihilated by the operators $\{\hat{a}_l\}$, $\{\hat{b}_l\}$ and the $\hat{b}_l$ modes are in the vacuum) and apply Eq.\ (\ref{unitary1}) to it, we find using the Baker-Campbell-Hausdorff formula, that 
\begin{align}
    \hat{U}\hat{a}_\text{in}^\dd | 0 \rangle &= \bigg( \cos \theta \sum_l f_l^\star \hat{a}_l^\dd - i \sin \theta \sum_l f_l^\star \hat{b}_l^\dd \bigg) | 0 \rangle.
\end{align}
This is the well-known input-output relationship for the interaction of $\hat{a}_\text{in}$ and $\hat{b}_\text{in}$ at a \textit{passive beamsplitter}, and is true for any choice of $f_l$. Note especially that Eq.\ (\ref{unitary1}) affects all of the constituent modes in $\hat{a}_\text{in}$ and $\hat{b}_\text{in}$ unilaterally, and cannot be considered mode-selective. 

To understand the action of a mode-selective mirror on the input modes, let us decompose $\hat{a}_\text{in}$ as
 \cite{Rohde_2007}
\begin{align}
    \hat{a}_\text{in} &= \big[ \hat{a}_\text{in},\hat{a}_0^\dd \big]\hat{a}_0  + \sum_{l\neq 0} f_l \hat{a}_l
\end{align}
where $\big|\big[\hat{a}_\text{in},\hat{a}_0^\dd \big]\big|^2 + \sum_{l\neq 0} |f_l|^2 = 1$ enforces normalisation. One approach is to use the following unitary:
\begin{align}\label{3}
    \hat{U}_0 &= \exp \bigg( -i \theta \bigg( \hat{a}_0 \hat{b}_0^\dd + \text{H.c} \bigg) \bigg).
\end{align}
Equation (\ref{3}) takes a similar form to the passive beamsplitter unitary of Eq.\ (\ref{unitary1}), however the key difference is that from the complete sets $\{\hat{a}_l\}$ and $\{\hat{b}_l\}$, it only affects $\hat{a}_0$ and $\hat{b}_0$. To see this, let us again prepare a single-photon state $\hat{a}_\text{in}^\dd | 0 \rangle$ and apply the unitary of Eq.\ (\ref{3}) to it, yielding
\begin{align}\label{3.2}
    \hat{U}_0\hat{a}_\text{in}^\dd | 0 \rangle &= \Big( \big[ \hat{a}_\text{in}, \hat{a}_0^\dd \big]^\star \big( \hat{a}_0^\dd \cos\theta - i \hat{b}_0^\dd \sin\theta \big) + \sum_{l\neq 0 } f_l^\star \hat{a}_l^\dd \Big) | 0 \rangle 
\end{align}
where we have utilised the properties $\hat{U}_0\hat{U}_0^\dd = \mathds{I}$ and $\hat{U}_0 |0 \rangle = | 0 \rangle$. 
If we consider $\theta = \pi/2$ (perfect reflection), we obtain
\begin{align}
    \hat{U}_0 \hat{a}_\text{in}^\dd | 0 \rangle &= \Big( - i  [\hat{a}_\text{in}, \hat{a}_0^\dd]^\star \hat{b}_0^\dd + \sum_{l\neq  0 } f_l^\star \hat{a}_l^\dd \Big) | 0 \rangle .
\end{align}
The overlap of $\hat{a}_0$ and $\hat{a}_\text{in}$ has been reflected into the $\hat{b}_0$ mode, while the orthogonal parts of the mode, $\hat{a}_{l\neq 0}$, are unaffected by the mirror. Thus, Eq.\ (\ref{3.2}) reflects a single mode (in the `0' mode) from the multi-mode input, while the orthogonal modes $\hat{a}_l$ pass through the mirror unaffected. The desired outcome is achieved, namely the isolation of a single mode from the input field.

\section{Continuous-variable teleportation as mode-selectivity}\label{sec:CVII}
In this section, we introduce the homodyne-measurement and all-optical approaches to continuous-variable (CV) teleportation of an input mode. This simple model highlights the mode-selective property of such teleporters, wherein a single mode from the input is transmitted through the circuit while all orthogonal modes are filtered or reflected. This is complementary to the unitary considered in Eq.\ (\ref{3}), where the selected mode was reflected while the orthogonal modes were transmitted. For our purposes, the two protocols achieve an equivalent result.  

\subsection{Homodyne measurement telefilter}\label{IIItelefiltersec}
Quantum teleportation is the process whereby an unknown input state, $\hat{\rho}_\text{in}$, is transferred between distant observers using a classical channel and a preexisting entanglement resource. The seminal work by Bennett et.\ al.\ \cite{BennettPhysRevLett.70.1895} utilised entangled Bell pairs as the resource, however this was later extended to continuous-variable protocols by Vaidman \cite{vaidmanPhysRevA.49.1473} and Braunstein and Kimble \cite{BraunsteinPhysRevLett.80.869}. Essentially, a bipartite entangled system is shared between a sender and receiver, who perform local operations and communicate via classical channels so that the receiver can retrieve an arbitrarily good version of the initial state $\hat{\rho}_\text{in}$, without the direct transmission of quantum information between them. 

\begin{figure}[h]
    \centering
    \includegraphics[width=0.85\linewidth]{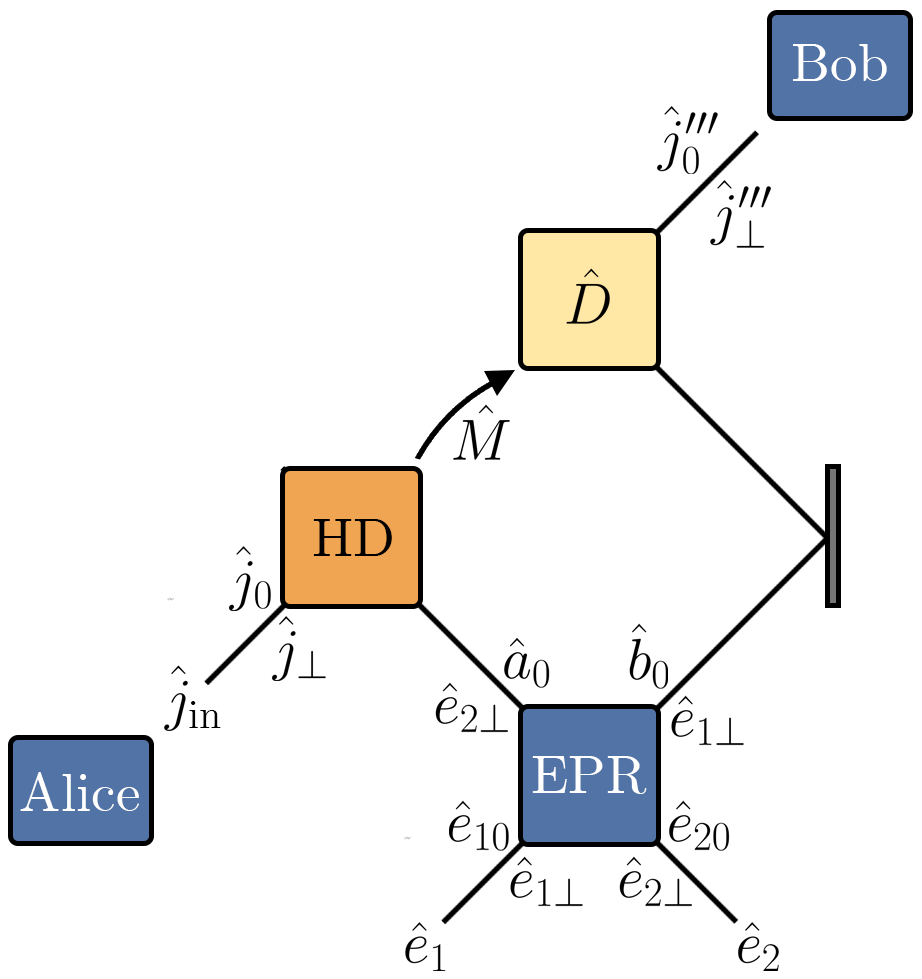}
    \caption{Schematic diagram of CV teleportation using a homodyne measurement of the input mode $\hat{j}_\text{in}$ with the entanglement resource mode $\hat{a}_0$. The grey mirror on the right is perfectly reflecting and simply included for aesthetic purposes; the left-moving complement mode incident on the other side is irrelevant to the protocol. The details of the homodyne measurement are illustrated in Fig.\ \ref{fig:homodynedetail}.}
    \label{fig:homodyne1mode}
\end{figure}

\begin{figure}[h]
    \centering
    \includegraphics[width=0.525\linewidth]{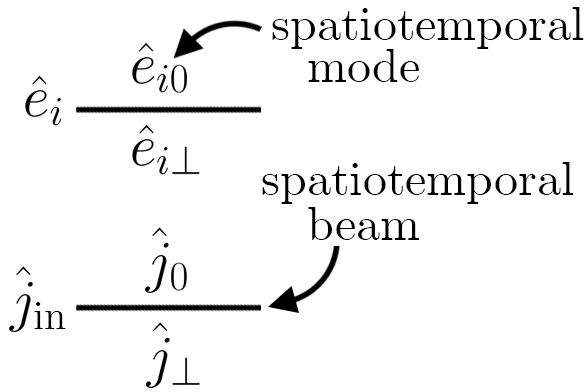}
    \caption{Schematic representation of the mode decomposition of the spatiotemporal modes, $\hat{e}_i$ and  $\hat{j}_\text{in}$. The orthogonal modes $\{\hat{e}_{i0},\hat{e}_{i\perp} \}$ and $\{\hat{j}_0,\hat{j}_\perp\}$ respectively co-propagate in the same spatiotemporal beam, denoted by the black line. }
    \label{fig:beamsplit}
\end{figure}

Throughout this paper, we will describe our quantum-optical protocols in the Heisenberg picture \cite{bachor2004guide}. The first approach is illustrated in the circuit diagram in Fig.\ \ref{fig:homodyne1mode}, which utilises a distributed entanglement resource and a dual homodyne measurement to achieve the mode-selective teleporatation. Firstly, Alice receives an input mode $\hat{j}_\text{in}$ in an arbitrary state. As discussed previously, $\hat{j}_\text{in}$ can be decomposed in the basis of constituent, orthonormal modes $\{\hat{j}_0,\hat{j}_\perp\}$ which co-propagate in the same spatiotemporal beam. This decomposition is illustrated schematically in Fig.\ \ref{fig:beamsplit}. She aims to select -- that is, transmit -- $\hat{j}_0$ (more precisely, the overlap of $\hat{j}_\text{in}$ with $\hat{j}_0$) to Bob, while filtering out $\hat{j}_\perp$. 

To achieve this, Alice enacts the dual homodyne measurement of $\hat{j}_0$ mixed with her half of the entanglement resource mode, $\hat{a}_0$. The entanglement resource modes are generated by two mode-sqeezing of the vacuum modes $\hat{e}_{i}$. Just like Alice's input mode, the input vacua can be decomposed in the orthonormal basis $\{\hat{e}_{i0},\hat{e}_{i\perp}\}$; these modes likewise co-propagate in the same beam. The $\hat{e}_{i0}$ modes are spatiotemporally mode-matched with the mode of interest, $\hat{j}_0$.
A two-mode squeezer, $\hat{S}_2(s)$, generates the two-mode squeezed state from the $\hat{e}_{i0}$ modes as follows:
\begin{align}\label{entanglement1}
    \hat{a}_0 = \hat{S}_2^\dd(s) \hat{e}_{10} \hat{S}_2(s) &= \cosh (s) \hat{e}_{10} + \sinh (s) \hat{e}_{20}^\dd \vt \\ \label{entanglement2}
    \hat{b}_0 = \hat{S}_2^\dd(s) \hat{e}_{20} \hat{S}_2(s) &= \cosh (s) \hat{e}_{20} + \sinh (s) \hat{e}_{10}^\dd , \vt 
\end{align}
where
\begin{align}
    \hat{S}_2(s) = \exp \Big( \xi^\star \hat{e}_{10} \hat{e}_{20} - \xi^\star \hat{e}_{10}^\dd \hat{e}_{20}^\dd \Big)
\end{align}
and $\xi = s e^{i\varphi}$, $\cosh(s)$ is the squeezing gain and $\varphi$ the phase. In the limit $s \to \infty$, $\hat{a}_0$ and $\hat{b}_0$ become perfectly entangled. The two-mode squeezing interaction could be mediated by a $\chi^{(2)}$ nonlinear crystal in which the shape of the pump field is mode-matched to the $\hat{e}_{i0}$ modes. We note here that the squeezer is mode-discriminating (not mode-selective); while it selectively generates a two-mode squeezed state of the $\hat{e}_{i0}$ vacuum modes, the `$\perp$' modes are still transmitted through the optical element without being affected. One recalls that our definition of mode-discrimination is to be distinguished from mode-selectivity, the latter being the desired property of the mirror where only the mode of interest is \textit{transmitted}, while the orthogonal modes are filtered or reflected. The entanglement resource modes, $\hat{a}_0$ and $\hat{b}_0$, are distributed to the two participants, Alice and Bob. 

\begin{figure}[h]
    \centering
    \includegraphics[width=0.9\linewidth]{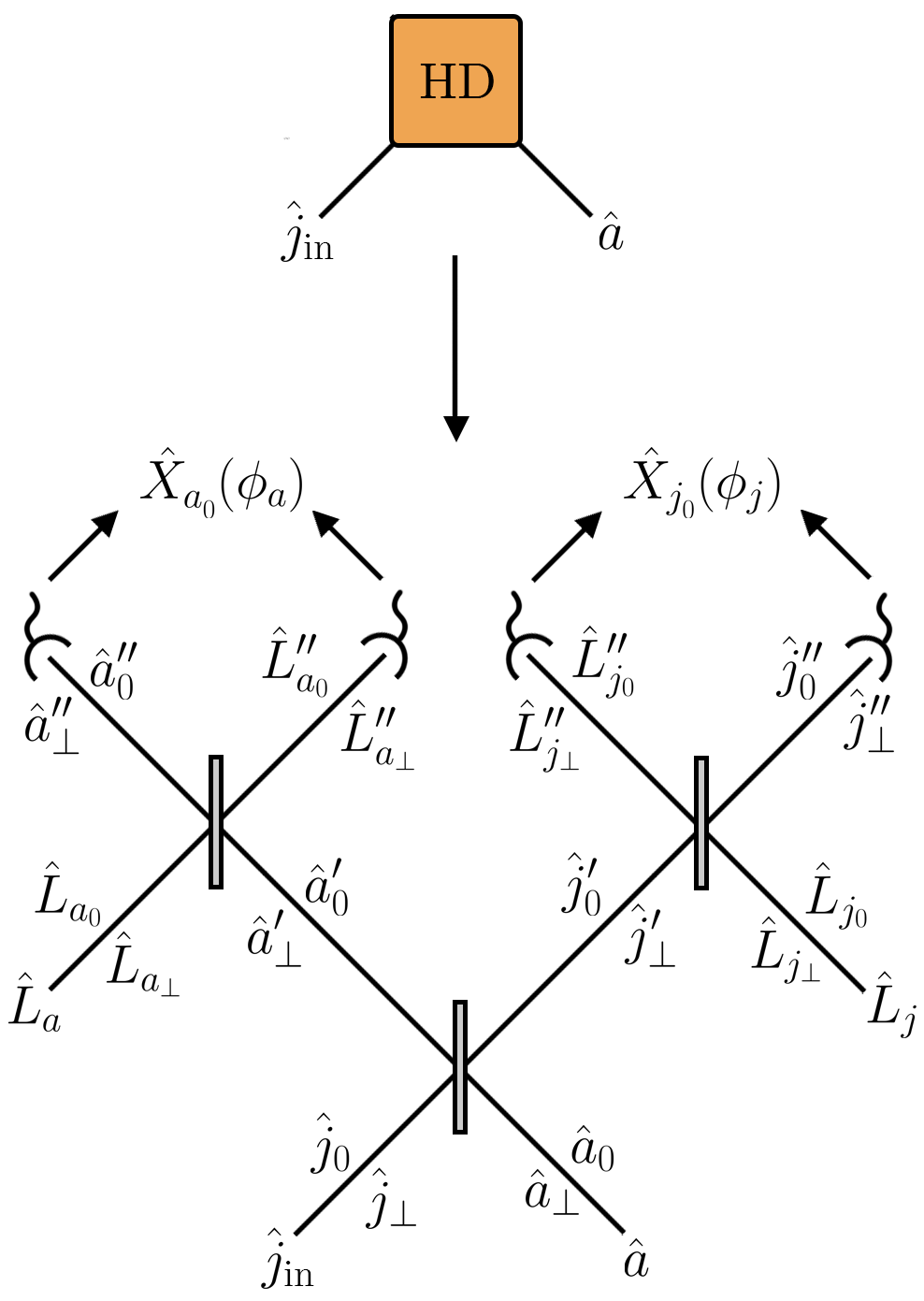}
    \caption{Circuit diagram of the homodyne detection scheme. $\hat{j}_\text{in}$ and $\hat{a}$ are mixed on a balanced beamsplitter. The measurement results are used to construct the operator $\hat{M}$.}
    \label{fig:homodynedetail}
\end{figure}

Let us return to Alice and her input mode, $\hat{j}_\text{in}$, which can be expressed as a superposition of the constituent orthonormal basis modes, 
\begin{align}\label{13}
    \hat{j}_\text{in} &= f_0 \hat{j}_0 + \sum_{l\neq 0 } f_l \hat{j}_l
\end{align}
alongside its orthogonal complement, $\hat{j}_\text{in,$\perp$}$,
\begin{align}\label{14}
    \hat{j}_\text{in,$\perp$}  &= \sqrt{1 - |f_0|^2} \hat{j}_0 - \frac{f_0}{\sqrt{1 - |f_0|^2}} \sum_{l\neq 0 } f_l \hat{j}_l 
\end{align}
where $\sum_l |f_l| = 1$. By making the association $\sqrt{\epsilon}  = f_0$, Eq.\ (\ref{13}) and (\ref{14}) can be recast into a beamsplitter type relation:
\begin{align}\label{152}
    \hat{j}_\text{in} &= \sqrt{\epsilon} \hat{j}_0 + \sqrt{1 - \epsilon} \hat{j}_\perp \vt \\ \label{162}
    \hat{j}_\text{in,$\perp$} &= \sqrt{1 - \epsilon} \hat{j}_0 - \sqrt{\epsilon} \hat{j}_\perp  \vt 
\end{align}
where we have defined
\begin{align}
    \hat{j}_\perp &= \frac{1}{\sqrt{1- | f_0|^2}} \sum_{l\neq 0 } f_l \hat{j}_l. 
\end{align}
Inverting Eq.\ (\ref{152}) and (\ref{162}), we obtain the following form for the constituent modes:
\begin{align}
    \hat{j}_0 &= \sqrt{\epsilon} \hat{j}_\text{in} + \sqrt{1 - \epsilon} \hat{j}_\text{in,$\perp$} \vt \\
    \hat{j}_\perp &= \sqrt{1 - \epsilon} \hat{j}_\text{in}  - \sqrt{\epsilon} \hat{j}_\text{in,$\perp$} . \vt 
\end{align}
% In a similar way, the entanglement resource modes, $\hat{a}$ and $\hat{b}$, can be likewise decomposed:
% \begin{align}
%     \hat{a}_0 &= \sqrt{\epsilon} \hat{a} + \sqrt{1- \epsilon} \hat{a}_\perp \\
%     \hat{a}_{0,\perp} &= \sqrt{1- \epsilon} \hat{a} - \sqrt{\epsilon} \hat{a}_\perp \\
%     \hat{b}_0 &= \sqrt{\epsilon} \hat{b} + \sqrt{1 - \epsilon} \hat{b}_\perp \\
%     \hat{b}_{0,\perp} &= \sqrt{1- \epsilon } \hat{b} - \sqrt{\epsilon} \hat{b}_\perp 
% \end{align}
Alice's dual homodyne measurement (a CV equivalent of a Bell measurement) on the mode of interest mixed with $\hat{a}_0$, is illustrated in Fig.\ \ref{fig:homodynedetail}.
Firstly, she mixes $\hat{j}_\text{in}$ and $\hat{a}$ (i.e.\ their constituent modes) at a balanced, passive beamsplitter, yielding
\begin{align}\label{20inputs}
\begin{split}
    \hat{j}_0' &= \frac{1}{\sqrt{2}} \big( \hat{a}_0 + \hat{j}_0 \big) ,
    \\
    \hat{a}_0' &= \frac{1}{\sqrt{2}} \big( \hat{j}_0 - \hat{a}_0 \big) ,
\end{split}
\begin{split}
    \hat{j}_\perp' &= \frac{1}{\sqrt{2}} \big( \hat{a}_\perp + \hat{j}_\perp \big), \\
    \hat{a}_\perp' &= \frac{1}{\sqrt{2}} \big( \hat{j}_\perp - \hat{a}_\perp \big) .
\end{split}
\end{align}
Alice prepares two local oscillator modes, $\hat{L}_j$ and $\hat{L}_a$, which can be similarly represented in a complete basis of orthogonal modes, $\{\hat{L}_{j_0}$, $\hat{L}_{j_\perp}\}$ and $\{\hat{L}_{a_0}$, $\hat{L}_{a_\perp}\}$ respectively. The subscripts denote the modes which the local oscillators are matched to. The local oscillators matching the `0' modes are assumed to be prepared in large amplitude coherent states, written in the form
\begin{align}\label{24l}
    \hat{L}_i &= \beta + \delta \hat{L}_i \qquad \text{where $i = j_0, a_0$}
\end{align}
where $\beta = |\beta| e^{-i\phi_i} = \langle \hat{L}_i \rangle \gg 1$ and $\delta \hat{L}_i = \hat{L}_i - \beta$, while the orthogonal local oscillator modes are vacuum modes given by
\begin{align}
    \hat{L}_k &= \delta \hat{L}_k \qquad \text{where $k= j_\perp, a_\perp$}.
\end{align}
We have utilised the nomenclature $\delta\hat{L}_k$ to emphasise that these modes have zero mean. Alice interacts the input modes, Eq.\ (\ref{20inputs}), with the respective local oscillator modes at balanced, passive beamsplitters, yielding
\begin{align}\label{23homodyne}
\begin{split}
    \hat{j}_0'' &= \frac{1}{\sqrt{2}} \big( \hat{j}_0' + \hat{L}_{j_0} \big), \\
    \hat{L}_{j_0}'' &= \frac{1}{\sqrt{2}} \big( \hat{L}_{j_0} - \hat{j}_0 ' \big), 
\end{split}
\begin{split}
    \hat{a}_0'' &= \frac{1}{\sqrt{2}} \big( \hat{a}_0' + \hat{L}_{a_0} \big) ,\\
    \hat{L}_{a_0}'' &= \frac{1}{\sqrt{2}} \big( \hat{L}_{a_0} - \hat{a}_0' \big) ,
\end{split}
\end{align}
and likewise, 
\begin{align}\label{24homodyne}
\begin{split}
    \hat{j}_\perp'' &= \frac{1}{\sqrt{2}} \big( \hat{j}_\perp' + \hat{L}_{j_\perp} \big), \\
    \hat{L}_{j_\perp}'' &= \frac{1}{\sqrt{2}} \big( \hat{L}_{j_\perp} - \hat{j}_\perp' \big), 
\end{split}
\begin{split}
    \hat{a}_\perp'' &= \frac{1}{\sqrt{2}} \big( \hat{a}_\perp' + \hat{L}_{a_\perp} \big) ,\\
    \hat{L}_{a_\perp}'' &= \frac{1}{\sqrt{2}} \big( \hat{L}_{a_\perp} - \hat{a}_\perp' \big) .
\end{split}
\end{align}
Using the modes in Eq.\ (\ref{23homodyne}) and (\ref{24homodyne}), we firstly construct number operators at the respective detectors, 
\begin{align}\begin{split}
    \hat{N}_k &= \hat{k}^\dd \hat{k} \qquad \text{where $k = a_0''$, $a_\perp''$, $j_0''$, $j_\perp''$} \vt \\
    \hat{N}_l &= \hat{l}^\dd \hat{l} \qquad \text{where $l = L_{a_0}''$, $L_{a_\perp}''$, $L_{j_0}''$, $L_{j_\perp}''$}  \vt 
\end{split}
\end{align}
which are then added together as follows 
\begin{align}
\begin{split}
    \hat{N}_a &= \hat{N}_{a_0''} + \hat{N}_{a_\perp''}, \vt \\
    \hat{N}_{L_a} &= \hat{N}_{L_{a_0}''} + \hat{N}_{L_{a_\perp}''}, \vt 
\end{split}
\begin{split}
    \hat{N}_{j} &= \hat{N}_{j_0''} + \hat{N}_{j_\perp''} , \vt \\
    \hat{N}_{L_j} &= \hat{N}_{L_{j_0}''} + \hat{N}_{L_{j_\perp}''} . \vt 
\end{split}
\end{align}
The photon numbers at the respective ports are subtracted to obtain the output signal operators, $\hat{O}_j = \hat{N}_k - \hat{N}_{L_j}$ and $\hat{O}_a = \hat{N}_a - \hat{N}_{L_a}$, yielding 
\begin{align}
    \hat{O}_j &= \hat{j}_0'^\dd \hat{L}_{j_0} + \hat{L}_{j_0}^\dd \hat{j}_0' + \hat{j}_\perp'^\dd \hat{L}_{j_\perp} + \hat{L}_{j_\perp}^\dd \hat{j}_\perp' \vt \\
    \hat{O}_a &= \hat{a}_0'^\dd \hat{L}_{a_0} + \hat{L}_{a_0} \hat{a}_0' + \hat{a}_\perp'^\dd \hat{L}_{a_\perp } + \hat{L}_{a_\perp}^\dd \hat{a}_\perp'. \vt 
\end{align}
By applying the definition of Eq.\ (\ref{24l}) and neglecting terms not multiplied by the coherent signal $\beta$, we find that 
\begin{align}
     \hat{O}_{j} &= |\beta| \hat{X}_{j_0'}(\phi_j) = | \beta| \big( e^{-i\phi_j} \hat{j}_0' + e^{i\phi_j} \hat{j}_0'^\dd \big) \vt   \\
   \hat{O}_{a} &= |\beta| \hat{X}_{a_0'}(\phi_a) = | \beta| \big( e^{-i\phi_a} \hat{a}_0' + e^{i\phi_a} \hat{a}_0'^\dd \big) , \vt 
\end{align}
where $\hat{X}_{j_0}(\phi_j)$, $\hat{X}_{a_0}(\phi_a)$ are the phase-dependent quadrature operators of the modes denoted in the subscript \cite{bachor2004guide}. These operators represent the classical statistics of the measurement. By taking $\phi_a$ and $\phi_j$ to be $\pi/2$ out-of-phase, Alice can subsequently construct 
an operator $\hat{M}$ from this joint measurement, for example
\begin{align}\label{measurementchannel}
    \hat{M} &= |\beta| (\hat{X}_{j_0'} + i \hat{P}_{a_0'}) = \sqrt{2}|\beta| \big( \hat{j}_0 - \hat{a}_0^\dd  \big)
\end{align}
which satisfies $[ \hat{M} , \hat{M}^\dd ] = 0$ and we have adopted the nomenclature
\begin{align}
    \hat{X}_i &= \hat{X}_i(\phi_i = 0 ) , \qquad \hat{P}_i = \hat{X}_i(\phi_i = \pi/2) . \vt     
\end{align}
$\hat{M}$ represents the classical channel which Alice sends to Bob \cite{ralphPhysRevA.65.012319}, who subsequently uses it to displace his half of the entangled pair, 
\begin{align}\label{hommeasurement11}
    \hat{j}_0''' &= \hat{b}_0 + \zeta \hat{M}
\end{align}
where $\zeta\in \mathbb{C}$ is the effective gain of the classical channel. Taking $\zeta = 1/(\sqrt{2}|\beta|)$ yields a unity gain channel between Alice and Bob, so that Bob's mode becomes
\begin{align}
    \hat{j}_0''' &= \hat{j}_0 + \big(\hat{b}_0 - \hat{a}_0^\dd\big). 
\end{align}
Using Eq.\ (\ref{entanglement1}) and (\ref{entanglement2}), we find that in the limit of perfect entanglement ($s \to \infty$), 
\begin{align}
    \lim_{s \to \infty}\hat{j}_0''' &= \hat{j}_0 , \qquad \hat{j}_\perp''' = \ \hat{e}_{1\perp},
\end{align}
that is, Bob perfectly retrieves the overlap of Alice's mode with the mode of interest, $\hat{j}_0$. Notice that the orthogonal mode $\hat{j}_\perp$ is filtered out at the measurement while the co-propagating mode $\hat{e}_{1\perp}$ is in the vacuum.

To reiterate why the teleporter is mode-selective, recall that only the mode of interest, $\hat{j}_0$, is matched to and consequently amplified by the large-amplitude local oscillator mode. Likewise, the entanglement resource $\hat{a}_0$ is prepared in the same spatiotemporal mode as $\hat{j}_0$ and $\hat{L}_{a_0}$. The sets of orthogonal modes $\{\hat{j}_\perp\}$ and $\{\hat{a}_\perp\}$ are filtered out at the level of the detection process, since they are not mixed with the large amplitude local oscillator modes. Thus, their unamplified quadratures are hidden beneath the semiclassical noise of the large-amplitude coherent state. In view of this, we refer to the circuit of Fig.\ \ref{fig:homodyne1mode} as a mode-selective telefilter. This is a complementary scenario to Eq.\ (\ref{3.2}), where the mode of interest was reflected while all others were transmitted. For our purposes, they achieve an identical outcome in terms of the isolation of the desired mode. 

\subsection{Telefilters with imperfect efficiency}\label{sec:imperfectefficiency}
The previous model for the single-mode telefilter was characterised by a unity gain channel between Alice and Bob; $\hat{j}_0$ is teleported between them without loss, and the efficiency of the protocol is effectively characterised by the additional noise carried by the entanglement resource modes. In the limit of infinite squeezing, the noise from these modes vanishes.

It is instructive to consider a non-unity gain channel between Alice and Bob, when the squeezing is finite. Such a scenario represents a telefilter with imperfect efficiency; for finite squeezing, some of the input mode is actually lost. Recall that Bob's output displaced mode is given by Eq.\ (\ref{hommeasurement11}), 
\begin{align}
    \hat{j}_0''' &= \hat{b}_0 + \zeta \hat{M}.
\end{align}
By taking the effective gain to be the squeezing-dependent function, $\zeta = \tanh(s)/(\sqrt{2}|\beta|)$ \cite{Ralph_1999}, we find that the output mode reduces to
\begin{align}\label{eq24inefficient}
    \hat{j}_0''' &= \tanh(s) \hat{j}_0 + \frac{\hat{e}_{20}}{\cosh(s)}
\end{align}
Since $\tanh(s)$ is a monotonically increasing function of the squeezing parameter $s$ within the range $[0,1]$, this determines how much of the input mode is successfully teleported (i.e.\ transmitted through the filter). In the limit of infinite squeezing, the coefficient of $\hat{e}_{20}$ vanishes, leaving $\hat{j}_0$ exactly. In this sense, the telefilter may be understood as a filter with a variable transmission coefficient determined by the squeezing parameter, $s$. As we show in Appendix \ref{appendix:imperfectelemirror}, this calculation generalises to an $N$-mode input.

\subsection{All-optical telemirror}\label{allopticaltelemirroratemporal}
\begin{figure}[h]
    \centering
    \includegraphics[width=0.725\linewidth]{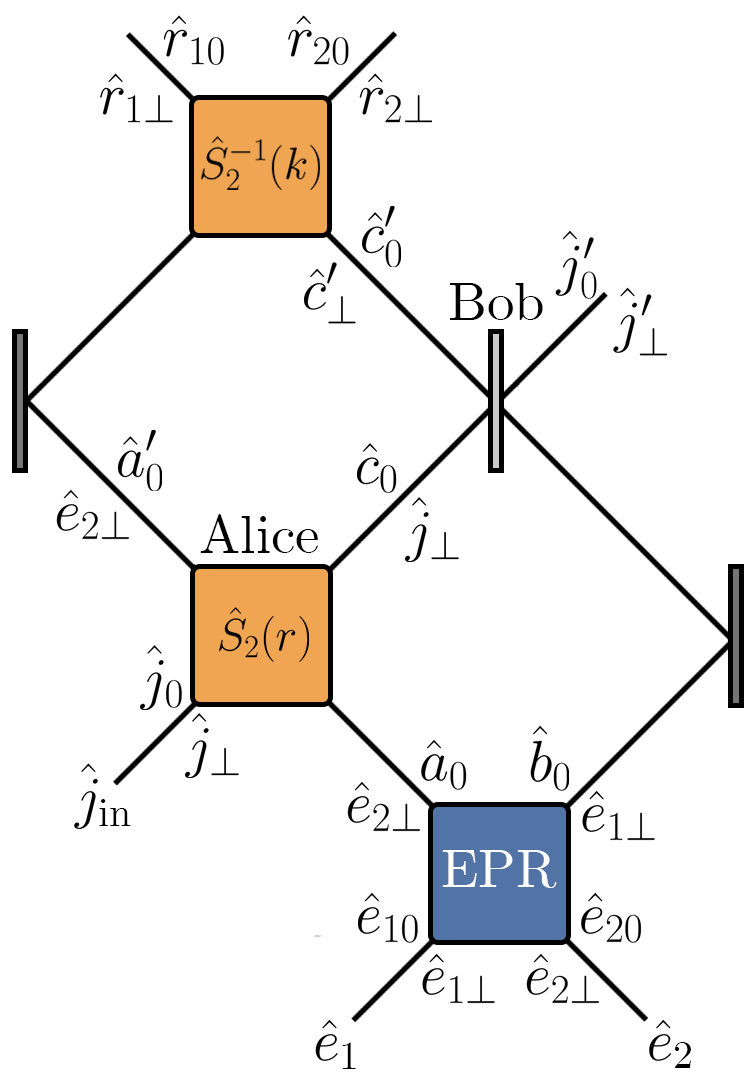}
    \caption{Schematic diagram of the all-optical teleportation circuit. The homodyne measurement is replaced by a two-mode squeezer, denoted $\hat{S}_2(r)$.}
    \label{fig:allopticalsingle}
\end{figure}

An alternative approach to CV teleportation is the all-optical protocol first proposed by Ralph in \cite{Ralph:99}. The difference between the all-optical and homodyne measurement protocols is that the former does not directly enact a measurement on the input mode, and is thus entirely unitary. Apart from this difference, the two approaches are functionally equivalent in terms of the teleported beam. Notably, the all-optical protocol has been recently realised in experiment \cite{liu2020orbital} and has also been recently studied in the context of relativistic, noninertial observers \cite{FooPhysRevD.101.085006}. The circuit diagram for the protocol is shown in Fig.\ \ref{fig:allopticalsingle}.
It can be straightforwardly shown that this is an equivalent approach to the previous case studied, using homodyne measurements. Unlike the telefilter, which utilises a dual homodyne measurement of the incoming modes, the input mode of interest, $\hat{j}_0$, is mixed with the entangled mode, $\hat{a}_0$, at a two-mode squeezer. The two-mode squeezer creates the classical channel between Alice and Bob by highly amplifying the input mode $\hat{j}_0$ along with $\hat{a}_0$, producing the mode $\hat{c}_0$. The quadratures of $\hat{c}_0$ have uncertainty significantly larger than the quantum noise limit of unity. Any noise introduced by a joint measurement of $\hat{X}_{c_0}$ and $\hat{P}_{c_0}$ is negligible compared to these already amplified amplitudes, warranting the designation of $\hat{c}_0$ as a classical field. 

This is where the mode-selectivity of the all-optical teleporter becomes manifest. As already mentioned, the two-mode squeezing interaction can be controlled by the shape of the pump field inside a $\chi^{(2)}$ nonlinear crystal, mode-matched with the mode to-be-squeezed. Here we assume that $\hat{j}_0$ and $\hat{a}_0$ are mode-matched with the mode-discriminating squeezer, while the orthogonal modes $\hat{j}_\perp$ and $\hat{a}_\perp$ pass through the crystal unaffected.

Now, the transformation which $\hat{S}_2(r)$ enacts upon the modes is expressed as 
\begin{align}\label{squeezingchannel}
    \hat{c}_0 &= \cosh (r) \hat{j}_0 + \sinh (r) \hat{a}_0^\dd, \vt \\ \label{15}
    \hat{a}_0' &= \cosh (r) \hat{a}_0 + \sinh ( r) \hat{j}_0^\dd . \vt 
\end{align}
In the limit of high gain ($r \gg 1$) the operator $\hat{c}_0$ is identical to the measurement operator in Eq.\ (\ref{measurementchannel}) (modulo an arbitrarily chosen squeezing phase between $\hat{j}_0$ and $\hat{a}_0$; we have chosen a ($+$) between them as the convention for the main text of this paper) for a large coherent amplitude $|\beta| \gg 1$. This emphasises the direct mapping between the two approaches at each stage of the protocol. As mentioned, the orthogonal modes pass through the squeezer unaffected. Bob uses a passive beamsplitter to mix his part of the entanglement, $\hat{b}_0$, with the classical channel. The output takes the form
\begin{align}\label{40}
    \hat{j}_\text{0}' &= \sqrt{\eta} \hat{c}_0 - \sqrt{1 - \eta} \hat{b}_0, \vt  \\
    \hat{j}_\perp' &= \sqrt{\eta} \hat{c}_\perp - \sqrt{1 - \eta}\hat{b}_\perp .  \vt 
\end{align}
Again, notice that Eq.\ (\ref{40}) is analogous to Eq.\ (\ref{hommeasurement11}) up to some phase determined by the squeezing angle and the beamsplitter phase. By setting $\eta = \cosh^{-2} (r)$ (i.e.\ most of the channel is reflected), this becomes
\begin{align}\label{allopticaloutput}
    \hat{j}_\text{0}' &= \hat{j}_0 - \tanh (r) ( \hat{b}_0 - \hat{a}_0^\dd) , \vt \\
    \hat{j}_\perp' &= \frac{1}{\cosh (r)} \hat{j}_\perp - \tanh(r) \hat{b}_\perp . \vt 
\end{align} 
In the limit of high entanglement between the resource modes $(r\to \infty)$, the outputs reduce to 
\begin{align}
    \hat{j}_0' &= \hat{j}_0,  \qquad 
    \hat{j}_\perp' = - \hat{e}_{1\perp},
\end{align}
that is, Bob reconstructs the overlap of Alice's mode with the mode of interest, $\hat{j}_0$, with the vacuum mode $\hat{e}_{1\perp}$ co-propagating with it. We thus see that both the homodyne measurement and all-optical approaches yield the same desired outcome, namely the transmission of a selected mode of interest. Importantly the mode-selection achieved by the protocol maps the input to the output in the same spatiotemporal mode. In comparison with the mode-selective telefilter, this protocol is entirely unitary which makes it possible to retrieve the reflected mode on Alice's side as well. For this reason, we refer to the all-optical teleporter as a mode-selective \textit{telemirror}. 

To determine the form of the reflected modes, one can perform an inverse squeezing operation, denoted $\hat{S}_2^{-1}(k)$, to retrieve the input vacua, $\hat{e}_{10}$ and $\hat{e}_{20}$ (see Fig.\ \ref{fig:allopticalsingle} and Appendix \ref{appendix1}). After taking the limit of infinite squeezing on both the entanglement and the classical channel, we obtain
\begin{align}\label{46}
\begin{split}
    \lim_{r,s \to \infty} \hat{r}_{10} &= \hat{e}_{10} ,\\
    \lim_{r,s\to \infty} \hat{r}_{20} &= \hat{e}_{20} ,
\end{split}
\begin{split}
    \lim_{r,s\to\infty} \hat{r}_{1\perp } &= \hat{j}_\perp, \\
    \lim_{r,s\to \infty } \hat{r}_{2\perp } &= \hat{e}_{2\perp} , 
\end{split}
\end{align}
which confirms that the teleportation protocol is fully unitary. The $\hat{j}_\perp$ mode orthogonal to $\hat{j}_0$ is reflected by the telemirror, since it does not interact with $\hat{S}_2(r)$. Finally, we note that as with the telefilter protocol, a similar calculation can be performed to obtain a telemirror with imperfect efficiency, which we present in Appendix \ref{appendix:imperfectelemirror}.

\section{Causality considerations in mode-selective interactions}\label{sec4causality}
\begin{figure}[h]
    \centering
    \includegraphics[width=0.65\linewidth]{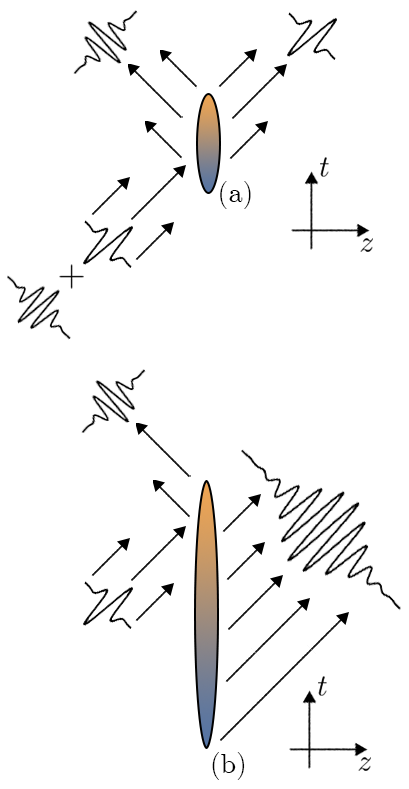}
    \caption{Schematic representation of the causality issues arising from the action of mode-selective mirrors. (a) represents a scenario similar to the mode-selective telemirror, where we have ignored the temporal aspects of the problem. In (b), the action of the mirror occurs over some delocalised time-window, which allows for superluminal signalling if instantaneous measurements are permitted. In particular, the transmitted mode has non-zero probability of the photon appearing in regions that are spacelike separated from the region in which the input pulse is located. }
    \label{fig:causalityviolation}
\end{figure}

In the previous sections, we reviewed two well-known CV teleportation protocols. By decomposing the input mode into co-propagating orthogonal components, we showed that such protocols act 
as mode-selective devices, transmitting the mode of interest $\hat{j}_0$ and filtering or reflecting its orthogonal complement mode, $\hat{j}_\perp$. However in these protocols, and more generally 
mode-selective interactions of the form of Eq.\ (\ref{3}), the temporal properties of the input mode are neglected, as are any causal considerations affecting the propagation of such modes between Alice and Bob. This scenario is illustrated schematically in Fig.\ \ref{fig:causalityviolation}(a). For this reason, we henceforth refer to the teleportation protocols discussed in Sec.\ \ref{sec:CVII} as \textit{atemporal} telefilters and telemirrors. 

Our aim for the remainder of this paper is to generalise our mode-selective teleportation model so that it captures the causality effects involved in the propagation of temporal modes \cite{tituPhysRev.145.1041,BrechtPhysRevX.5.041017,Ansari:18,Raymer_2020} through such mode-selective devices. Consider Fig.\ \ref{fig:causalityviolation}(b), in which the input mode only partially overlaps with the mirror. Since the temporal extent of the mode which the mirror selects is longer than that of the input mode, one can arrive at causality-violating scenarios in which superluminal signalling is possible if the mode-selective mirror acts instantaneously. This can be considered a quantum-optical iteration of Sorkin's well-known `impossible measurement' problem \cite{Sorkin:1993gg}. We ask: what is the consistent treatment for the propagation of temporally extended modes through such mirrors which obeys relativistic and causality considerations?

\section{Telefilters and telemirrors with temporal mode inputs}\label{timedelay}
In Sec.\ \ref{sec:II}, we showed how continuous-variable teleportation protocols function as mode-selective telefilters and telemirrors which act on an input multi-mode field. The mode of interest, $\hat{j}_0$, is selected by the telefilter (or telemirror) at the level of the homodyne measurement (or the generation of the classical channel), allowing it to propagate to Bob while all other modes are filtered or reflected. In the limit of a perfect entanglement resource, $\hat{j}_0$ is perfectly reconstructed by Bob.

In the following, we study a series of new teleportation protocols where the input field mode is distributed over two temporal modes. Using this approach allows us to explicitly investigate the action of the telemirror on an input which is delocalised in time. This can be extended to an $N$-mode input (Appendix \ref{sec:nmode}), which becomes a better approximation to a continuous spatiotemporal wavepacket as $N\to \infty$. The desired operation of this temporal mode teleporter is that a single superposition of the inputs (say, the symmetric superposition) is transmitted to the receiver, while the orthogonal superposition (the anti-symmetric superposition) and all other orthogonal mode components (previously denoted by the $\hat{j}_\perp$ modes) are filtered out or reflected. 

\subsection{Time-delayed temporal mode telefilter}\label{sec:time-delaytemporalmodetelefilter}
\begin{figure}[h]
    \centering
    \includegraphics[width=0.9\linewidth]{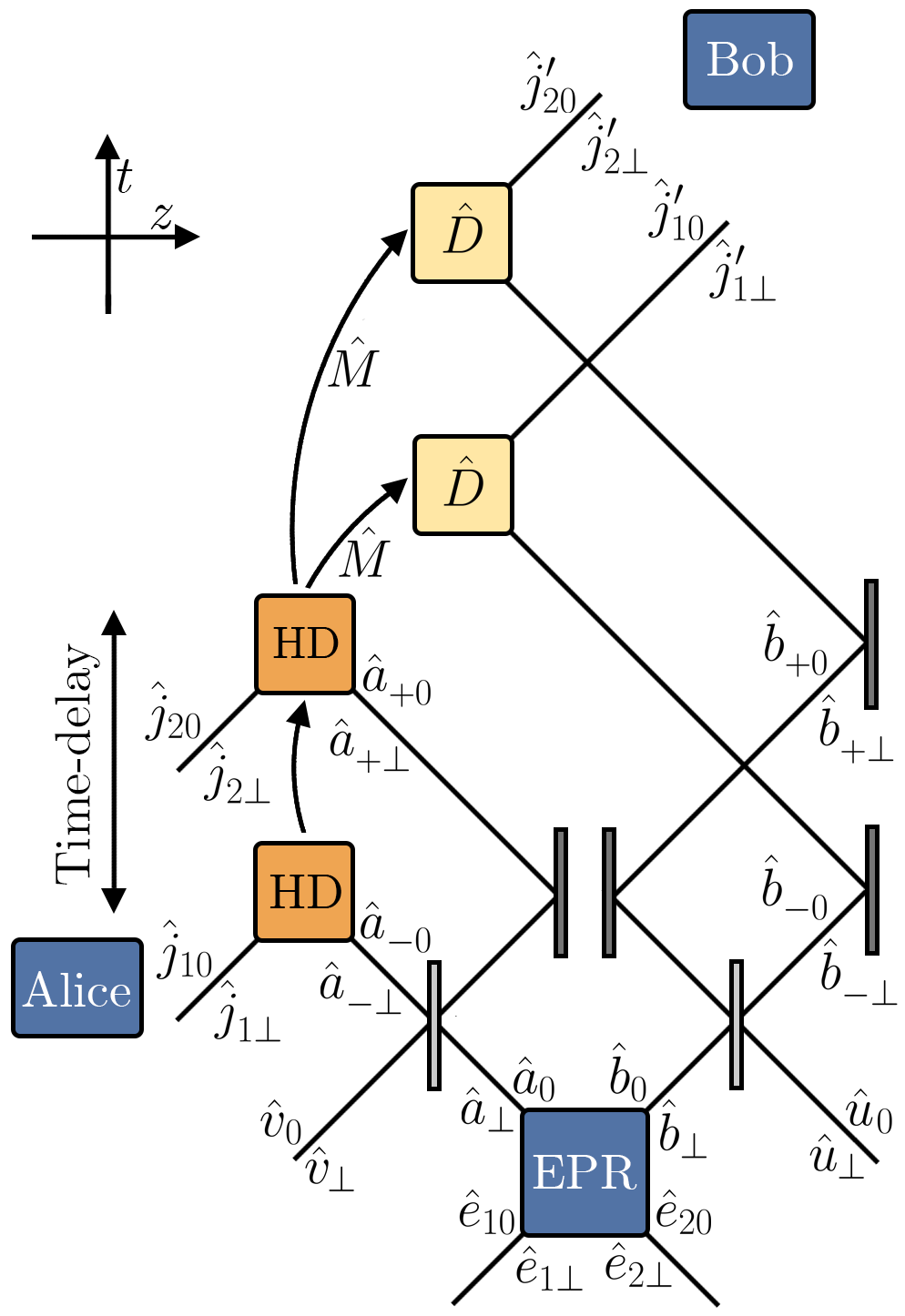}
    \caption{Circuit diagram for the temporal mode telefilter, which takes as inputs the temporal modes $\hat{j}_1$ and $\hat{j}_2$. The arrow connecting the two homodyne detectors signifies that the measurement results of $\hat{j}_{i0}$ are added together before displacing the $\hat{b}_{\pm0} $ modes respectively. The light grey mirrors are characterised by the transmission coefficient $\alpha$ and phase $\phi$. The time-delay between the second homodyne measurement and the displacements of $\hat{b}_{\pm0} $ can be made arbitrarily small; however, the fundamental time-delay on the mode propagation is constrained by the arrival times of the inputs $\hat{j}_1$ and $\hat{j}_2$.}
    \label{fig:homodyne2mode}
\end{figure}

We firstly consider what we refer to as the temporal mode telefilter, building off the atemporal mode-selective telefilter introduced in Sec.\ \ref{sec:II}. Figure \ref{fig:homodyne2mode} shows the new circuit diagram. 
Like the previous analyses, on each rail of the circuit the `$\perp$' modes co-propagate with the `0' modes, which are the set of modes which we are interested in selecting. Here, our assumption is that the local oscillator modes are prepared in the same spatiotemporal mode as $\hat{j}_{10}$ and $\hat{j}_{20}$, as well as the entanglement resource modes $\hat{a}_0$ and $\hat{b}_0$. 

Let us explain the protocol enacted by the temporal mode telefilter. Just as in Fig.\ \ref{fig:homodyne1mode}, the entangled beams $\hat{a}_0$ and $\hat{b}_0$ are sent to Alice and Bob respectively. Since the input is distributed over two temporal modes, the entanglement resource needs to be distributed likewise. It is the choice of the entanglement mode superposition which characterises the mode-selectivity of the temporal mode telefilter. The entanglement resource modes are distributed into two temporal modes using passive beamsplitters with transmission coefficient $\alpha$ and phase $\phi$:
\begin{align}
\begin{split} 
    \hat{a}_{-0}  &= \sqrt{\alpha}  - i e^{-i\phi} \sqrt{1 - \alpha} \hat{a}_0  \vt \\
    \hat{a}_{+0}  &= \sqrt{\alpha} \hat{a}_0 - i e^{i\phi} \sqrt{1 - \alpha} \hat{v}_0   \vt \\
    \hat{b}_{-0}  &= \sqrt{\alpha}  - ie^{-i\phi} \sqrt{1 - \alpha} \hat{b}_0 ,  \vt \\
    \hat{b}_{+0}  &= \sqrt{\alpha} \hat{b}_0 - i e^{i\phi}\sqrt{1 - \alpha} \hat{u}_0  . \label{22} \vt 
\end{split} 
\end{align}
To perform the teleportation protocol, we perform the dual homodyne measurements explicated in Sec.\ \ref{IIItelefiltersec} using the inputs, $\hat{j}_{10}$ and $\hat{j}_{20}$, mixed with the entanglement resource modes, $\hat{a}_{\pm0} $. This is achieved firstly by mixing $\hat{j}_{10}$ and $\hat{j}_{20}$ with the earlier and later parts of the entanglement resource modes respectively, 
\begin{align}
\begin{split} 
    \hat{j}_{10}' &=  \frac{1}{\sqrt{2}} (\hat{j}_{10} + \sqrt{\alpha} \hat{v}_0 - i e^{-i\phi } \sqrt{1 - \alpha}\hat{a}_0) \\
    \hat{a}_{-0} &= \frac{1}{\sqrt{2}} ( \hat{j}_{10} - \sqrt{\alpha} \hat{v}_0 + i e^{-i\phi } \sqrt{1 - \alpha}\hat{a}_0 ) \\
    \hat{j}_{20}' &= \frac{1}{\sqrt{2}} ( \hat{j}_{20} + \sqrt{\alpha} \hat{a}_0 - i e^{i\phi } \sqrt{1 - \alpha} \hat{v}_0 ) \\
    \hat{a}_{+0} &= \frac{1}{\sqrt{2}} ( \hat{j}_{20} - \sqrt{\alpha}\hat{a}_0 + i e^{i\phi } \sqrt{1 - \alpha}\hat{v}_0  )
\end{split} 
\end{align}
before coupling them with the coherently displaced local oscillator modes, detecting the photon numbers at the output ports and constructing the conjugate quadrature operators. These quadrature operators are then used to construct the appropriate measurement operators (i.e.\ the classical measurement result) given by:
\begin{align}
\begin{split}
    \hat{M}_1 &= \hat{X}_{\hat{j}_{10}'} + i\hat{P}_{\hat{a}_{-0}}, \vt \\
    \hat{M}_2 &=  \hat{X}_{\hat{j}_{20}'} +  i\hat{P}_{\hat{a}_{+0}}, \vt 
\end{split}
\begin{split}
    \hat{M}_1 = \hat{X}_{\hat{a}_{-0}} + i\hat{P}_{\hat{j}_{10}'} , \vt \\
    \hat{M}_2 = \hat{X}_{\hat{a}_{+0}} + i \hat{P}_{\hat{j}_{20}'} .\vt 
\end{split}
\end{align}
Now, recall that one may interpret $\hat{j}_{10}$ and $\hat{j}_{20}$ as the individual temporal components of a single input wavepacket mode to-be-selected by the teleporter. To successfully enact the teleportation of this `wavepacket', one has to combine the measurement results from each of the temporal modes, then use this to displace Bob's side of the entanglement resource, $\hat{b}_{\pm 0 } $. Clearly, one cannot perform a measurement of $\hat{j}_{20}$ before it has entered the apparatus; neither can the entire mode be teleported before $\hat{j}_{20}$ has been measured. This constraint time-orders the two independent measurement processes in the circuit, preventing $\hat{j}_{20}$ from being teleported before $\hat{j}_{10}$. To emphasise this point, one should consider Fig.\ \ref{fig:homodyne2mode} as occurring on a flat background with spacetime coordinates ($t,z$) (as shown at the top of the figure) where the rails in the circuit now represent the lightlike propagation of the modes. 

Returning to the measurement, we can construct a \textit{total measurement operator}, $\hat{M}$, as a linear combination of the individual results,
\begin{align}\label{MeasurementTotal}
    \hat{M} &= \zeta_1(\alpha,\phi) \hat{M}_1 + \zeta_2(\alpha,\phi) \hat{M}_2 
\end{align}
where $\zeta_i(\alpha,\phi) \in \mathbb{C}$. This summation over the measurement results is indicated by the arrow connecting the two homodyne measurements in the circuit of Fig.\ \ref{fig:homodyne2mode}. The total measurement operator is communicated classically to Bob (similarly depicted as arrows), which he uses to displace his part of the distributed entanglement resource modes, $\hat{b}_{\pm0} $. He obtains the output modes
\begin{align}
    \hat{j}_{10}' &= \hat{b}_{-0}  + \lambda_1(\alpha,\phi) \hat{M}, \vt \\
    \hat{j}_{20}' &= \hat{b}_{+0}  + \lambda_2(\alpha,\phi) \hat{M} , \vt 
\end{align}
where $\lambda_i(\alpha,\phi) \in \mathbb{C}$. As we show in Appendix \ref{appendixarbitrary}, Alice can select any arbitrary superposition of the inputs $\hat{j}_{i0}$ by tuning the physical properties of the optical elements and the homodyne measurements.

Let us consider a simple case. In our construction, the phase of the quadrature operators in the measurement process, the distribution of the entanglement resource and the magnitude of the displacements made to Bob's side of the entanglement resource modes, completely determines the properties of the mode to-be-selected. The most straightforward example is obtained by taking the $\hat{X}$ and $\hat{P}$ quadratures at the homodyne measurement, and when $\hat{a}_{\pm 0} $ and $\hat{b}_{\pm 0} $ are distributed equally between the early and late temporal modes. In such a case, we should expect the transmitted mode to be a balanced superposition of the inputs $\hat{j}_{10}$ and $\hat{j}_{20}$. Taking for example $\alpha = 1/2$, $\phi = -\pi/2$ and the measurement results 
\begin{align}\label{59measurementchannel}
    \hat{M}_1 &= |\beta| (\hat{X}_{a_{-0}} + i \hat{P}_{j_{10}'}) = | \beta| ( \sqrt{2}\hat{j}_{10} - \hat{a}_0^\dd - \hat{v}_0^\dd ) \vt \\ \label{60measurementchannel}
    \hat{M}_2 &= |\beta| (\hat{X}_{a_{+0}} + i  \hat{P}_{j_{20}'}) = | \beta| ( \sqrt{2}\hat{j}_{20} - \hat{a}_0^\dd + \hat{v}_0^\dd )  \vt 
\end{align}
then we obtain for the total measurement operator
\begin{align}
    \hat{M} &=  \frac{1}{\sqrt{2}} ( \hat{j}_{10} + \hat{j}_{20} - \sqrt{2}\hat{a}_0^\dd ) 
\end{align}
having taken $\zeta_i(\alpha,\phi) = 1/(2\sqrt{2}|\beta|)$, which achieves a unity gain channel between Alice and Bob. Bob then displaces his part of the distributed entanglement, $\hat{b}_{\pm0} $, with the total measurement operator to obtain the outputs
\begin{align}
    \label{33b}
    \hat{j}_{10}' &= \frac{1}{2} \big( \hat{j}_{10} + \hat{j}_{20} \big) + \frac{1}{\sqrt{2}} \big( \hat{b}_0 - \hat{a}_0^\dd \big) + \frac{1}{\sqrt{2}}\hat{u}_0 , \\
    \hat{j}_{20}' &= \frac{1}{2} \big( \hat{j}_{10} + \hat{j}_{20} \big) + \frac{1}{\sqrt{2}} \big( \hat{b}_0 - \hat{a}_0^\dd \big) - \frac{1}{\sqrt{2}}  \hat{u}_0
\end{align}
where we have taken $\lambda_i(\alpha,\phi) = 1$, while the co-propagating `$\perp$' modes are $\hat{j}_{1\perp}' = \hat{b}_{-\perp} $ and $\hat{j}_{2\perp}' = \hat{b}_{+\perp} $. These can be mixed back together on a balanced beamsplitter to obtain the input vacuum modes:
\begin{align}
    \hat{o}_{1\perp}^{(l)} &= \frac{1}{\sqrt{2}} (\hat{j}_{2\perp}'  + \hat{j}_{1\perp}' ) = \hat{e}_{1\perp }, \\
    \hat{o}_{1\perp }^{(r)} &= \frac{1}{\sqrt{2}} ( \hat{j}_{1\perp }' - \hat{j}_{2\perp }' ) = \hat{u}_{1\perp }. 
\end{align}
Meanwhile for the `0' modes, we can likewise mix the outputs on a balanced beamsplitter, yielding
\begin{align}\label{42}
    \hat{o}_{10}^{(l)} &= \frac{1}{\sqrt{2}} \big( \hat{j}_{20}' + \hat{j}_{10}' \big) = \hat{u}_0 , \\ \label{36}
    \hat{o}_{10}^{(r)} &= \frac{1}{\sqrt{2}} \big( \hat{j}_{10}' - \hat{j}_{20}' \big)  = \frac{1}{\sqrt{2}} \big( \hat{j}_{10} + \hat{j}_{20} \big) + \big( \hat{b}_0 - \hat{a}_0^\dd \big) .
\end{align}
We find that the symmetric superposition of the input temporal modes has been successfully teleported to $\hat{o}_{10}^{(r)}$ along with the entanglement resource modes, while the unavoidable beamsplitter input $\hat{u}_0$ is isolated to the other output, $\hat{o}_{10}^{(l)}$. We conclude that the device successfully selects the mode of interest i.e.\ only one temporal superposition is transmitted, while the orthogonal superposition is filtered out at the level of the measurement. Any other modes which appear at the output are in the vacuum. In Appendix \ref{appendixarbitrary} we show how through an appropriate choice of the quadrature phases and the beamsplitter coefficients, Alice can select any arbitrary superposition mode to be transmitted to Bob.

Despite the success of our model in functioning mode-selectively, Fig.\ \ref{fig:homodyne2mode} also reveals that this mode-selectivity cannot be achieved without a time-delay which minimally matches the length of the input field mode. In our simplified two-mode case, the temporal length of the input `wavepacket' is simply the time-delay between modes $\hat{j}_{10}$ and $\hat{j}_{20}$. For teleportation to successfully occur, the individual classical measurement results enacted upon the input temporal modes must be coherently combined (added together) \textit{and then} used to displace Bob's side of the entanglement. Necessarily then, Alice must wait until the entire wavepacket (i.e.\ both the earlier and later temporal components) has been measured, before sending the classical channel to Bob. In one sense, one may be relieved that such a delay exists, meaning that the causality-violating scenario of Fig.\ \ref{fig:causalityviolation}(b) is not generally permitted. If one could enact the homodyne measurement of the entire wavepacket instantaneously, there would be no time-delay. Such a scenario is unphysical, and will inevitably lead to situations in which the input mode is teleported to an output mode which is earlier in time. 

\subsection{Time-delayed temporal mode telemirror}\label{allopticaltimedelay}
\begin{figure}[h]
    \centering
    \includegraphics[width=0.975\linewidth]{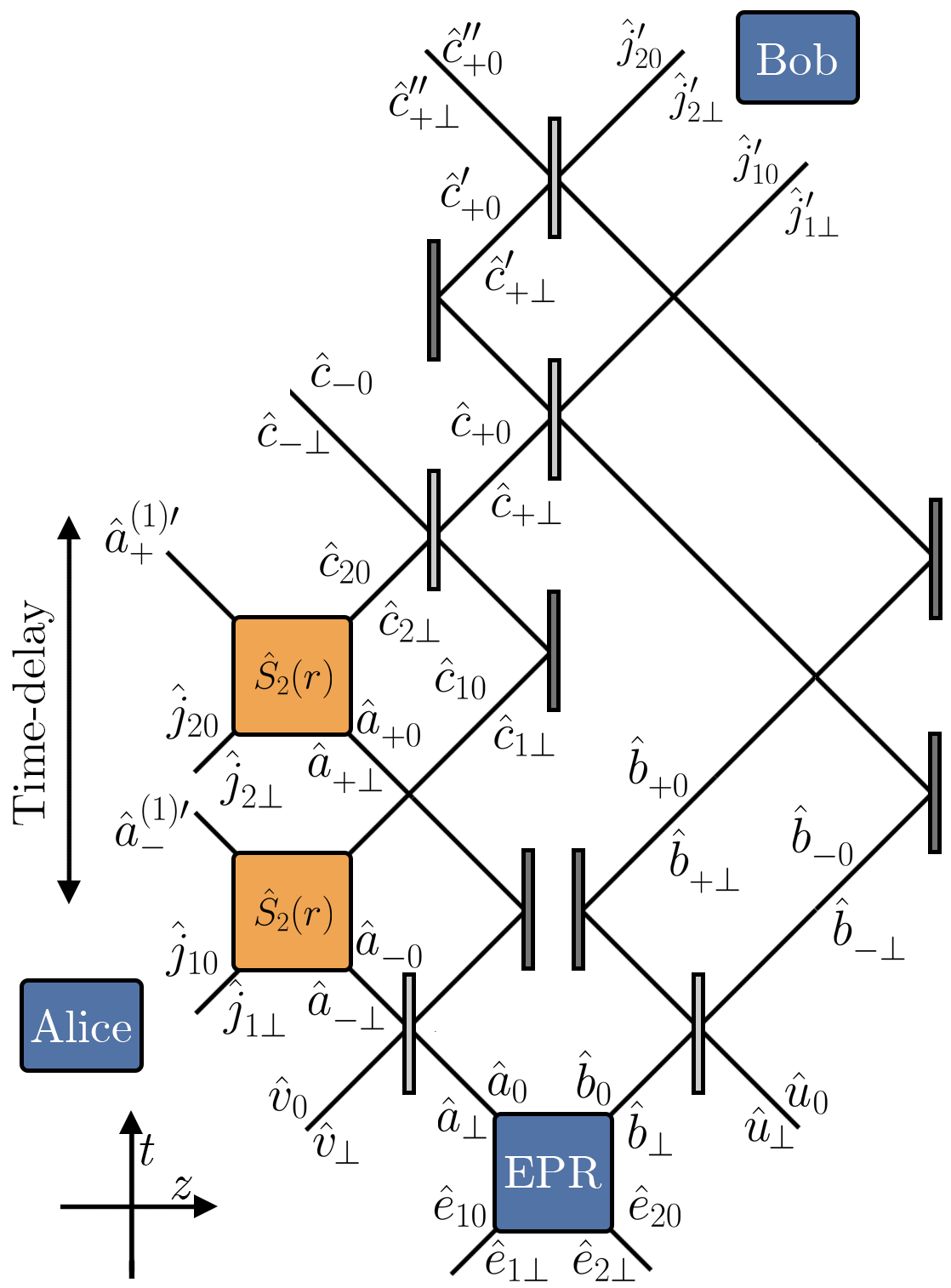}
    \caption{Circuit diagram for the temporal mode telemirror. The homodyne measurements are replaced by two-mode squeezing and the coherent displacements are replaced by beamsplitter interactions between the classical channel modes and the $\hat{b}_{\pm 0} $ modes. }
    \label{fig:timedelay2}
\end{figure}

We now consider the all-optical version of the mode-selective telemirror with two input temporal modes, shown schematically in Fig.\ \ref{fig:timedelay2}. This approach is fully unitary and also allows us to calculate both the transmitted and reflected modes. 
As with the temporal mode telefilter, the entanglement resource modes are distributed into spatial superpositions, Eq.\ (\ref{22}). Rather than mixing the entanglement with the input temporal modes and performing a homodyne measurement, the classical channel is enacted via two-mode squeezing,
\begin{align}\label{67}
    \hat{c}_{10} &= \cosh (r) \hat{j}_{10} + \sinh (r) \hat{a}_{-0}^{(1)\dd }, \vt  \\ \label{68}
    \hat{c}_{20} &= \cosh (r) \hat{j}_{20} + \sinh (r) \hat{a}_{+0}^{(1)\dd}, \vt 
\end{align}
while the other ports propagate to the reflected side:
\begin{align}
    \hat{a}_{-0} &= \cosh(r) \hat{a}_{-0}  + \sinh(r) \hat{j}_{10}^\dd , \vt  \\
    \hat{a}_{+0} &= \cosh(r) \hat{a}_{+0}  + \sinh(r) \hat{j}_{20}^\dd . \vt 
\end{align}
As with the atemporal versions of the telefilter and telemirror, Eq.\ (\ref{67}) and Eq.\ (\ref{68}) are identical to the measurement operators in Eq.\ (\ref{59measurementchannel}) and (\ref{60measurementchannel}) modulo some squeezing phase between $\hat{j}_{i0}$ and $\hat{a}_{\pm 0} $, in the limit of high gain (i.e.\ $r\gg 1$ for the telemirror and $|\beta| \gg 1$ in the telefilter). As usual, we assume that only the `0' modes and the associated entanglement resource modes are mode-matched to the squeezers generating the classical channel. The $\hat{c}_{i0}$ modes are highly amplified by the two-mode squeezers $\hat{S}_2(r)$ and can thus be regarded as good approximations to a classical field while the `$\perp$' modes remain unaffected by the squeezers.  

In the temporal mode telefilter, the individual measurement results are added together and used to displace the earlier and later components of Bob's entanglement resource mode, $\hat{b}_{\pm0} $. The equivalent operation in the all-optical version of the protocol is to mix the classical channel modes on a passive beamsplitter to obtain an analogous total measurement operator. We have
\begin{align}
    \hat{c}_{-0} &= \sqrt{\alpha} \hat{c}_{10} - i e^{i\phi_{c_0} } \sqrt{1 - \alpha}\hat{c}_{20} , \vt \\ 
    \hat{c}_{+0} &= \sqrt{\alpha}\hat{c}_{20} - i e^{-i\phi_{c_0} } \sqrt{1 - \alpha}\hat{c}_{10} . \vt 
\end{align}
The `$+$' part of this mode is likewise used to displace one of the temporal entanglement resource modes on Bob's side, 
\begin{align}
    \hat{j}_{10}' &= \sqrt{\eta_-} \hat{b}_{-0}  - i e^{-i\theta_- } \sqrt{1 - \eta_-} \hat{c}_{+0}, \vt  \\
    \hat{c}_{+0}' &= \sqrt{\eta_-} \hat{c}_{+0}  - i e^{i\theta_- } \sqrt{1 - \eta_-} \hat{b}_{-0}  , \vt 
\end{align}
while the leftover part of the channel is used to displace the later temporal entanglement resource mode:
\begin{align}
    \hat{j}_{20}' &= \sqrt{\eta_+} \hat{b}_{+0}  - i e^{- i\theta_+ } \sqrt{1 - \eta_+ }\hat{c}_{+0}', \vt \\
    \hat{c}_{+0}'' &= \sqrt{\eta_+} \hat{c}_{+0}'- i e^{i\theta_+ } \sqrt{1 - \eta_+} \hat{b}_{+0}  . \vt 
\end{align}
Let us show how to select the symmetric superposition of $\hat{j}_{i0}$. Firstly, let us take $ \phi = \phi_\pm = - \pi/2$ and $\phi_{c_0} = \pi/2$ with the beamsplitter transmission coefficients
\begin{align}
    \eta_\pm &= 1 - \frac{1}{2 \cosh^2(r)} .
\end{align}
We have chosen $\eta_\pm$ so that most of the highly amplified channel is reflected away from the output; in the limit of high gain, this ensures that the entanglement resource modes are transmitted to the output modes. The output modes (in the limit $r \to \infty$) are given by 
\begin{align}
    \hat{j}_{10}' &= - \frac{1}{2} ( \hat{j}_{10} + \hat{j}_{20} ) + \frac{1}{\sqrt{2}} ( \hat{b}_0 - \hat{a}_0^\dd ) + \frac{1}{\sqrt{2}} \hat{u}_0, \\
    \hat{j}_{20}' &= - \frac{1}{2} ( \hat{j}_{10} + \hat{j}_{20} ) + \frac{1}{\sqrt{2}} ( \hat{b}_{0 } - \hat{a}_0^\dd ) - \frac{1}{\sqrt{2}} \hat{u}_0.
\end{align}
The symmetric superposition of $\hat{j}_{i0}$ and the entanglement resource modes are transmitted to both output ports, along with an extra vacuum mode used to distribute the entanglement across two temporal modes. 
Mixing these finally on a balanced beamsplitter yields
\begin{align}
    \hat{o}_{10}^{(l)} &= \frac{1}{\sqrt{2}} ( \hat{j}_{20}'+ \hat{j}_{10}') = - \frac{1}{\sqrt{2}}( \hat{j}_{10} + \hat{j}_{20} ) + ( \hat{b}_0 - \hat{a}_0^\dd ) \\
    \hat{o}_{10}^{(r)} &= \frac{1}{\sqrt{2}} ( \hat{j}_{10}' - \hat{j}_{20}') = \hat{u}_0 \VT 
\end{align}
which is just the symmetric superposition of $\hat{j}_{10}$ and $\hat{j}_{20}$ modulo a global phase, carrying with it the entanglement resource modes, and vacuum in the other. This result is analogous to that found in Eq.\ (\ref{36}) where the mode of interest (the symmetric superposition, determined by the beamsplitter settings) is retrieved at one of the output ports, while the other mode merely contains the $\hat{u}_0$ mode coming from the beamsplitter mixing. This further illustrates how the telefilter and the unitary telemirror are equivalent in the transformations they enact on the selected mode as it is transmitted through the circuit. As we demonstrate in Appendix \ref{appendixarbitrary}, Alice can select any arbitrary superposition of $\hat{j}_{i0}$ by tuning the beamsplitter coefficients appropriately. 

We can also calculate the output reflected modes. Following a straightforward but tedious calculation in Appendix \ref{appendix1}, the following modes are reconstructed on the reflected side:
\begin{align}\label{74reflectedmode}
\begin{split}
    \hat{r}_{10} &= \hat{v}_0, \\
    \hat{r}_{20} &= \frac{1}{\sqrt{2}} ( \hat{j}_{10} - \hat{j}_{20} ) ,
\end{split}
\begin{split} 
    \hat{r}_{30} &= \hat{e}_{10},  \\
    \hat{r}_{40} &= \hat{e}_{20} \vphantom{\frac{1}{\sqrt{2}}}.
\end{split} 
\end{align}
We make two observations. Firstly, $\hat{r}_{10}$ contains the vacuum mode $\hat{u}_0$ used to prepare the temporal superposition of the entanglement resource modes, while $\hat{r}_{20}$ contains the antisymmetric superposition of the input temporal modes. This is the desired outcome; only the symmetric superposition is transmitted through the telemirror while its orthogonal complement, the anti-symmetric superposition, is reflected. Secondly, $\hat{r}_{30}$ and $\hat{r}_{40}$ contain the original vacuum modes used to generate the entanglement resource distributed to Alice and Bob. This demonstrates that the whole protocol is fully unitary. 

What about the other orthogonal modes? Recall that the `$\perp$' modes are unaffected by the two-mode squeezers $\hat{S}_2(r)$. By straightforwardly tracing the propagation of these modes through the circuit, we find that the `$\perp$' modes at Bob's side are given by $\hat{o}_{1\perp}^{(l)} = \hat{u}_{1\perp}$ and $\hat{o}_{1\perp }^{(r)} = \hat{e}_{1\perp}$ which are simply vacuum modes involved in the entanglement generation, while the reflected outputs are given by 
\begin{align}
\begin{split} 
    \hat{r}_{1\perp} &= \hat{e}_{2\perp }, \vt \\
    \hat{r}_{2\perp} &= \hat{v}_{1\perp}, \vt 
\end{split}
\begin{split} 
    \hat{r}_{3\perp} &= \hat{j}_{1\perp} , \vt  \\
    \hat{r}_{4\perp } &= \hat{j}_{2\perp}. \vt 
\end{split} 
\end{align}
The all-optical telemirror nicely illustrates the necessity of the time-delay as the input temporal modes pass through the circuit. From Fig.\ \ref{fig:timedelay2}, a fundamental time-delay is imposed on the teleportation of the temporal modes, since the channels must be combined to form the total measurement operator, $\hat{c}_{+0}$. This delay is a function of the temporal length of the mode itself; in our model, the time-delay between the arrival of the input pulses $\hat{j}_{10}$ and $\hat{j}_{20}$. 
As in the temporal mode telefilter, this constraint ensures the preservation of causality by time-ordering the unitary operators so that they act on the input modes in a consistent temporal order. This prevents the later input mode $\hat{j}_{20}$ from arriving at an output which is earlier in time.

\section{Mode-selective teleporters with no time-delay}\label{causal}
The temporal mode telefilter and telemirror demonstrate that in general, mode-selective quantum-optical mirrors enact an unavoidable and intrinsic time-delay on the propagation of input temporal modes. Such an effect was not observed in the original CV teleportation protocols because the temporal dimension of the input modes was neglected. This is akin to neglecting time-ordering effects in the original unitary for the mode-selective beamsplitter, Eq.\ (\ref{3}). 

Nevertheless, our results beg the question of what goes wrong if we try to construct a mode-selective mirror interaction which does not induce an intrinsic time-delay. Rather than integrating over the entire temporal length of the mode (e.g.\ waiting to combine the measurement results on the individual modes together), our next models consider the teleportation of the individual temporal modes separately and continuously before coherently recombining them at the output. We study two approaches: in the first, there are two independent entanglement resources for each of the temporal modes, while in the second, we distribute the entanglement resource modes into earlier and later components, as was done for the temporal mode teleporters introduced previously. For simplicity, we will just consider the selection of an equal symmetric superposition and the filtering or reflection of its anti-symmetric pair.

\subsection{Teleportation with independent entanglement resources}
\begin{figure}[h]
    \centering
    \includegraphics[width=\linewidth]{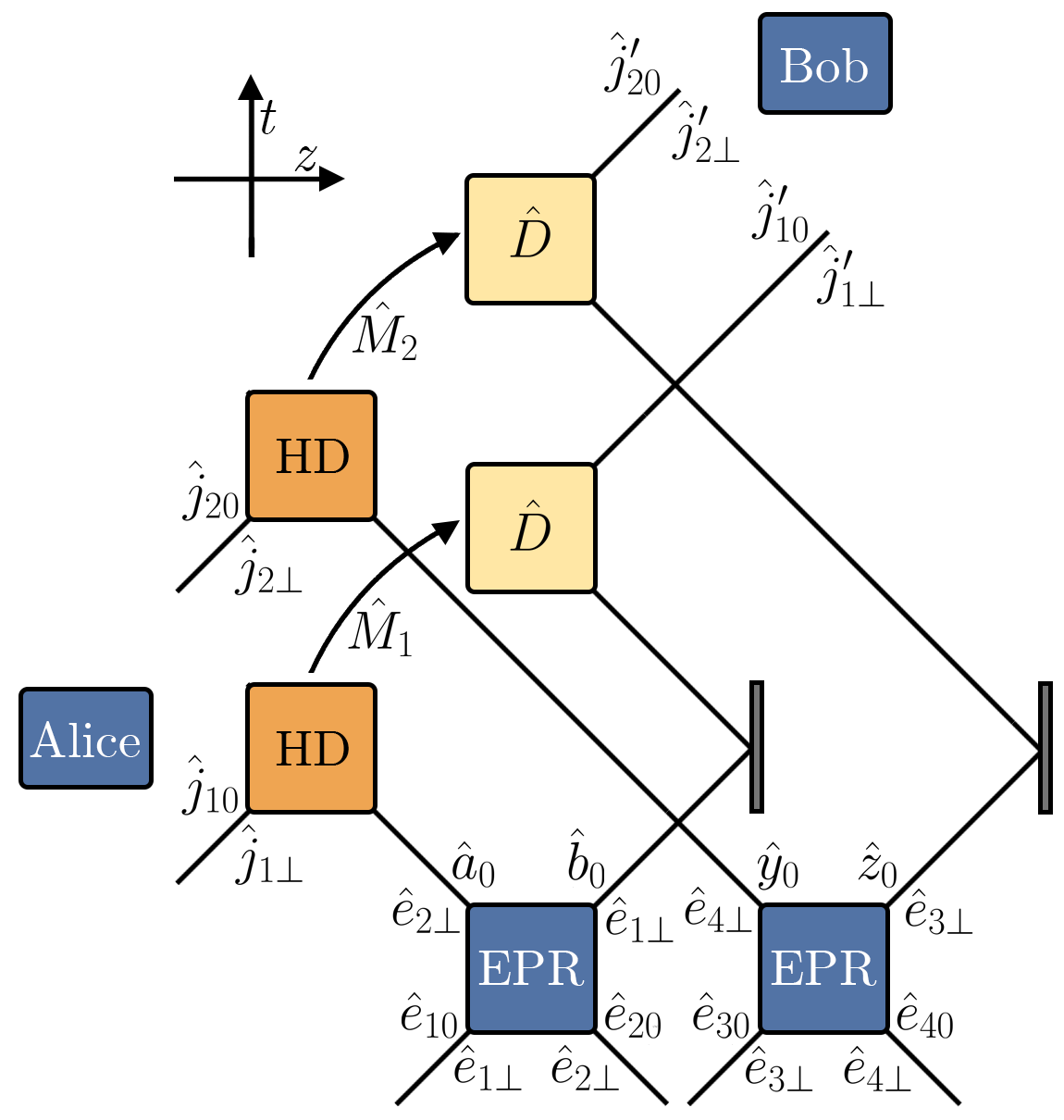}
    \caption{Circuit diagram for the no-delay telefilter with independent entanglement resources. $\hat{y}_0$ and $\hat{z}_0$ represent the entanglement resource modes used to teleport $\hat{j}_{20}$.}
    \label{nodelay0}
\end{figure}

Our first approach to circumventing the time-delay is shown in Fig.\ \ref{nodelay0}.
As before, we utilize homodyne measurements to obtain the results $\hat{M}_i$, which represent the classical channel connecting Alice and Bob. Alice performs the individual measurements separately and sends them continuously to Bob as she obtains them. This avoids the fundamental time-delay introduced in the temporal mode telefilter and telemirror. 

It is straightforwardly shown that the output of the circuit is identical to the homodyne measurement telefilter, but now teleports both temporal modes:
\begin{align}
\begin{split}
    \hat{j}_{10}' &= \hat{j}_{10} + (\hat{b}_{0} - \hat{a}_0^\dd ), \vt  \\
    \hat{j}_{20}' &= \hat{j}_{20} + (\hat{y}_0 - \hat{z}_0^\dd) , \vt 
\end{split}
\begin{split}
    \hat{j}_{1\perp}' &= \hat{b}_{\perp}, \vt  \\
    \hat{j}_{2\perp }' &= \hat{z}_\perp , \vt 
\end{split}
\end{align}
where $\hat{y}_0$ and $\hat{z}_0$ are generated by two-mode squeezing of independent vacuum inputs. In the limit of perfect entanglement between $\hat{b}_0$ and $\hat{a}_0$, $\hat{y}_0$ and $\hat{z}_0$, the $\hat{j}_{i0}$ modes are individually recovered at the outputs. This of course presents an immediate and obvious issue; any temporal superposition of $\hat{j}_{i0}$, in addition to its orthogonal complement, can be reconstructed by Bob. For example, 
\begin{align}
    \hat{o}_{10}^{(l)} &= \frac{1}{\sqrt{2}}( \hat{j}_{10}' + \hat{j}_{20}' ) = \frac{1}{\sqrt{2}} ( \hat{j}_{10} + \hat{j}_{20} ) \vt \\
    \hat{o}_{10}^{(r)} &= \frac{1}{\sqrt{2}} ( \hat{j}_{10}' - \hat{j}_{20}') = \frac{1}{\sqrt{2}} ( \hat{j}_{10} - \hat{j}_{20} ) . \vt 
\end{align}
Thus, while the time-delay on the propagation of the input modes is avoided, the introduction of two separate entanglement resource mode pairs means that the protocol ultimately fails to be mode-selective. Nor can it be considered mode-discriminating, because all superposition modes with temporal support on $\hat{j}_{i0}$ are teleported ideally. 

\subsection{No-delay temporal mode telefilter}\label{sec:nodelaytelefilter}
\begin{figure}[h]
    \centering
    \includegraphics[width=0.85\linewidth]{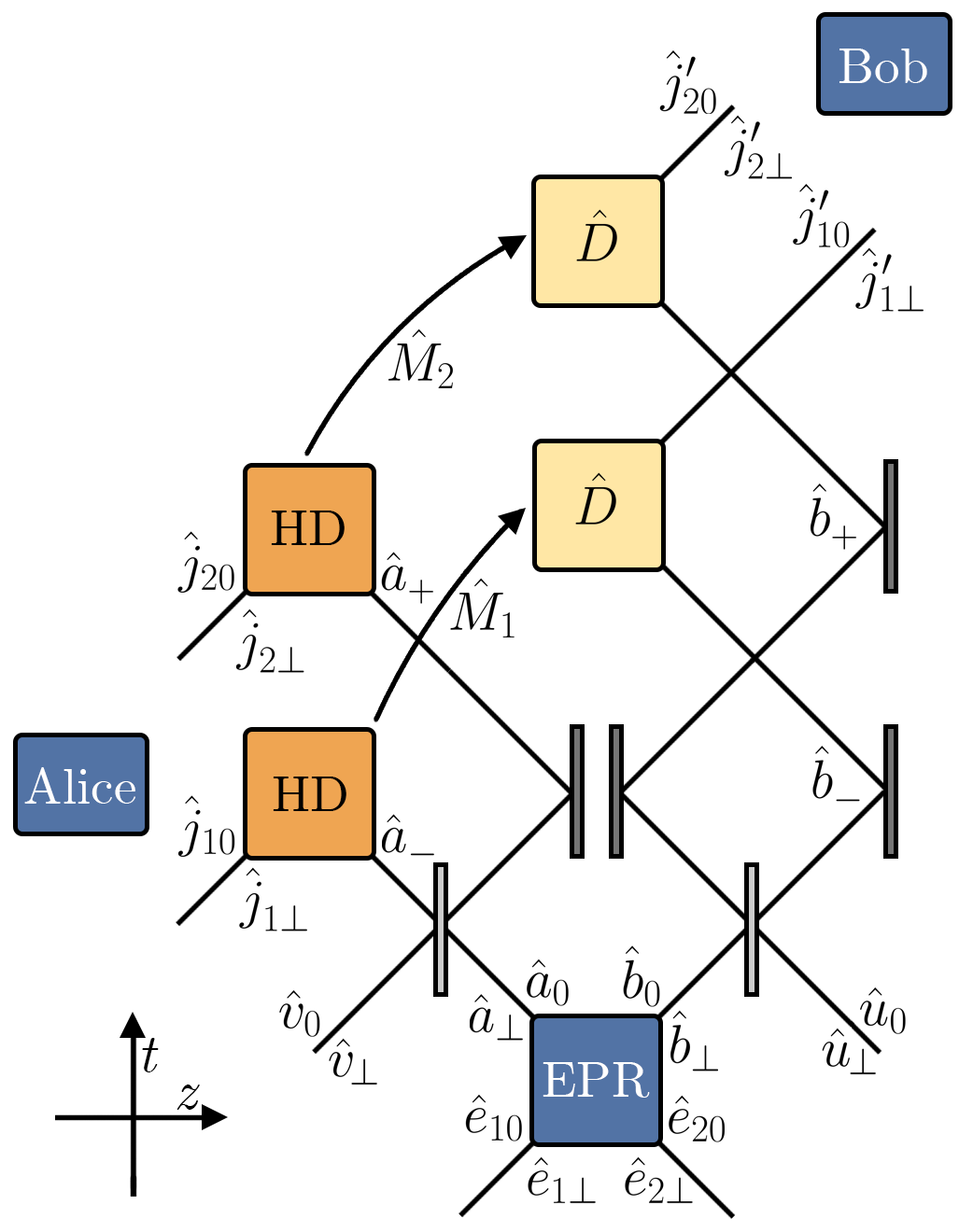}
    \caption{Circuit diagram for the no-delay telefilter. The $\hat{a}_{\pm} $ and $\hat{b}_\pm $ should be understood as containing the orthogonal co-propagating `0' and `$\perp$' modes.}
    \label{fig:homodyne2modecausal}
\end{figure}

Instead of utilizing independent entanglement resources, we now take the previous approach by distributing a single entanglement resource over two temporal modes. Our proposed circuit for this model, which we refer to as the no-delay telefilter, is displayed in Fig.\ \ref{fig:homodyne2modecausal}. 
As usual, homodyne measurements are enacted by mode-matching the entanglement resource mode with the input mode and local oscillator to obtain the classical results $\hat{M}_i$. Unlike the previous model, where Alice would add the measurement results of both modes together (requiring her to wait until both $\hat{j}_{10}$ and $\hat{j}_{20}$ had entered the apparatus), Alice sends the individual measurement results $\hat{M}_1$ and $\hat{M}_2$ to Bob `on-the-run'. 

Alice and Bob distribute the entanglement resource modes in a superposition of earlier and later components. Let us consider the same beamsplitter settings as Sec.\ \ref{sec:time-delaytemporalmodetelefilter}. Using again the measurement results
\begin{align}\label{85measurementchannel}
    \hat{M}_1 &=  |\beta| (\hat{X}_{a_{-0}} + i \hat{P}_{j_{10}'}) = | \beta| ( \sqrt{2}\hat{j}_{10} - \hat{a}_0^\dd - \hat{v}_0^\dd ), \vt  \\ \label{86measurementchannel}
    \hat{M}_2 &= |\beta| (\hat{X}_{a_{+0}} + i  \hat{P}_{j_{20}'}) = | \beta| ( \sqrt{2}\hat{j}_{20} - \hat{a}_0^\dd + \hat{v}_0^\dd ) , \vt 
\end{align}
Bob displaces the individual modes $\hat{b}_{\pm0} $ to obtain the following outputs:
\begin{align}
\hat{j}_{10}' &= \hat{j}_{10} + \frac{1}{\sqrt{2}} \big( \hat{b}_0 - \hat{a}_0^\dd \big) + \frac{1}{\sqrt{2}} \big( \hat{u}_0 - \hat{v}_0^\dd \big) \label{seventyfour}, \\
\hat{j}_{20}' \label{eq:72}
&= \hat{j}_{20} + \frac{1}{\sqrt{2}} \big( \hat{b}_0 - \hat{a}_0^\dd \big) - \frac{1}{\sqrt{2}} \big( \hat{u}_0 - \hat{v}_0^\dd \big) ,
\end{align}
where we have used $\zeta_i = 1/(\sqrt{2}|\beta|)$ to achieve a unity gain channel for the individual temporal components. As can be seen, the earlier and later input modes are teleported to the earlier and later output modes respectively, with two sources of additional noise from the beamsplitter inputs. Mixing the output modes back together on a balanced beamsplitter yields
\begin{align}\label{106}
    \hat{o}_{10}^{(l)} &= \frac{1}{\sqrt{2}} \big( \hat{j}_{10} + \hat{j}_{10} \big) + \big( \hat{b}_{0} - \hat{a}_{0}^\dd \big) , \\ \label{107}
    \hat{o}_{10}^{(r)} &=  \frac{1}{\sqrt{2}} \big( \hat{j}_{10} - \hat{j}_{20} \big) + \big( \hat{u}_0 - v_0^\dd \big) .
\end{align}
Similar to teleportation with independent entanglement resources, both the symmetric and anti-symmetric superpositions are transmitted to Bob. The difference is that the symmetric superposition has been teleported along with the entanglement resource modes while the anti-symmetric superposition has been teleported with additional noise from the beamsplitter inputs.
As such, this second model for the no-delay telefilter exhibits a kind of mode-discriminating property. Rather than uniquely affecting a single mode and transmitting all others through the identity channel, the converse is true; the selected mode is transmitted without change while the orthogonal mode is polluted by noise. Furthermore, in the limit where these extraneous modes are highly noisy, any signal encoded in the anti-symmetric superposition mode cannot be detected by Bob, as it is buried within this noise. This could be achieved if for example, $\hat{u}_0$ and $\hat{v}_0$ are generated from independent entanglement sources, where the other half of the entangled pairs are sent to the reflected side of the mirror. Bob, on the receiving side of the mirror, traces out these reflected modes and thus, cannot purify $\hat{o}_{10}^{(l)}$ to retrieve the orthogonal mode. As shown in Appendix \ref{appendixarbitrary}, one can tune the beamsplitter coefficients appropriately so that any specified superposition of the temporal modes $\hat{j}_{i0}$ can be transmitted along with the entanglement resource and thus be discriminated from its orthogonal superposition mode which carries the additional noise.
\begin{figure}[h]
    \centering
    \includegraphics[width=0.475\linewidth]{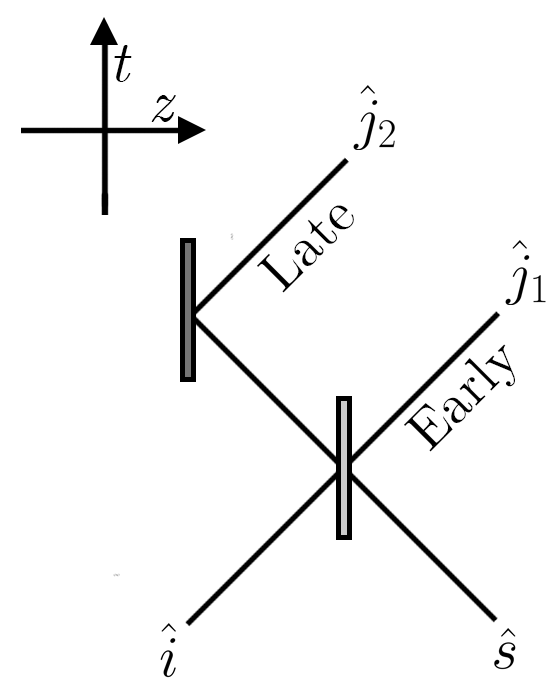}
    \caption{Schematic diagram of the input modes, split into a superposition of early and late time-bins, $\hat{j}_{10}$ and $\hat{j}_{20}$, at a beamsplitter. $\hat{i}$ could be prepared in a single-photon state while $\hat{s}$ is in the vacuum.}
    \label{fig:splitphoton}
\end{figure}

How does the no-delay telefilter prevent violations of causality? Let us prepare a mode, $\hat{i}$, in a single-photon state in a localised time-bin mode prior to the input temporal modes $\hat{j}_{10}$ and $\hat{j}_{20}$, see Fig.\ \ref{fig:splitphoton}. We note that this input could be in some arbitrary state, but using a single photon is conceptually straightforward. From Eq.\ (\ref{seventyfour}) and (\ref{eq:72}), one sees that if a photon is prepared in the `late' mode $\hat{j}_{20}$, it cannot be transmitted to the `early' mode, $\hat{j}_{10}'$. Neither can the converse be true; a photon prepared in the `early' mode $\hat{j}_{10}$ cannot be transmitted to the `late' mode, $\hat{j}_{20}'$. Causality is manifestly preserved.

What information about the temporal structure of the mode sent by Alice can Bob extract? If he attempts to make measurements in the symmetric superposition basis, he will detect the single photon half of the time, and vacuum for the rest. This is expected if the photon is initially prepared in either of the temporal modes. To perform such a measurement, this requires an additional time-delay (to recombine the outputs $\hat{j}_{i0}'$ together) so that the entire mode can be measured. Alternatively, Bob could attempt to measure the photon in \textit{either} the early or late output temporal modes, Eq.\ (\ref{seventyfour}) and Eq.\ (\ref{eq:72}). However such an attempt will be thwarted by the presence of unavoidable excess noise on both of the modes. As a result, Bob cannot distinguish whether the photon was in the early or late output temporal mode.

\subsection{No-delay temporal mode telemirror}\label{sec:nodelaytelemirror}
For completeness, in this section we provide a brief discussion of the no-delay temporal mode telemirror, see Fig.\ \ref{fig:causalteleportermain}. The full calculation of the input-output relations is shown in Appendix \ref{appendixarbitrary}. 
\begin{figure}[h]
    \centering
    \includegraphics[width=0.85\linewidth]{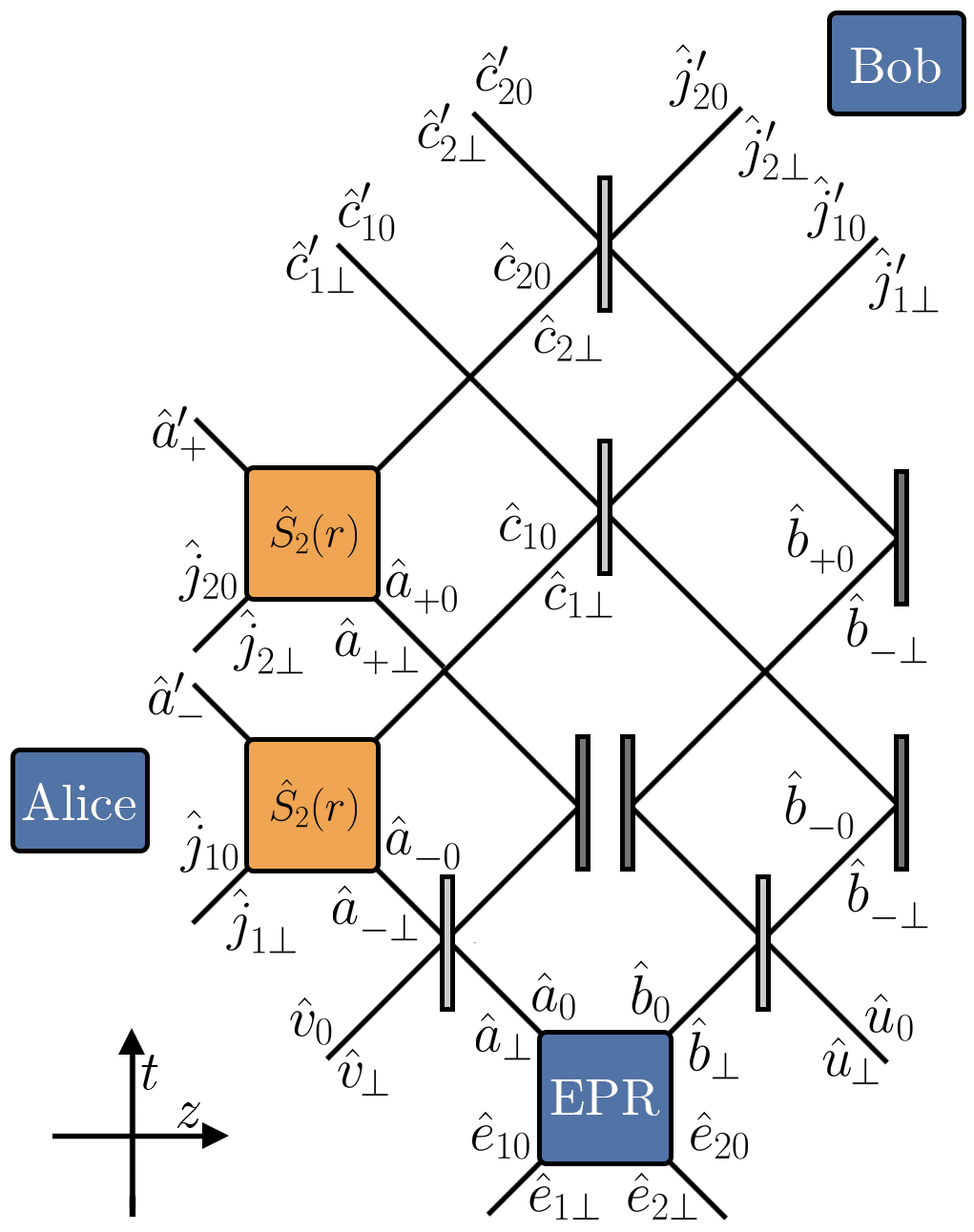}
    \caption{Circuit diagram of the no-delay temporal mode telemirror.}
    \label{fig:causalteleportermain}
\end{figure}
As usual, the entanglement resource is distributed into early and late temporal modes, which interact with the input temporal modes at two-mode squeezers, generating the classical channel. Like the temporal mode telefilter, the individual channels are mixed with Bob's half of the entanglement resource modes, from which he recovers as outputs (choosing an equal distribution of the entanglement resource, and appropriate beamsplitter phases),
\begin{align}\label{82main}
\hat{j}_{10}' &= \hat{j}_{10} - \frac{\tanh(r)}{\sqrt{2}} ( \hat{b}_0 - \hat{a}_0^\dd ) - \frac{\tanh(r)}{\sqrt{2}} ( \hat{u}_0 - \hat{v}_0^\dd ) ,\\ \label{83main}
\hat{j}_{20}' &= \hat{j}_{20} - \frac{\tanh(r)}{\sqrt{2}} ( \hat{b}_0 - \hat{a}_0^\dd ) + \frac{\tanh(r)}{\sqrt{2} } (\hat{u}_0 - \hat{v}_0^\dd ) .
\end{align}
Equations (\ref{82main}) and (\ref{83main}) are analogous to Eq.\ (\ref{seventyfour}) and (\ref{eq:72}) obtained for the no time-delay telefilter (modulo arbitrarily chosen phase factors). After mixing the modes together, one of the outputs perfectly teleports the symmetric superposition of $\hat{j}_{10}$ and $\hat{j}_{20}$, while the other teleports the anti-symmetric superposition classically, with the additional noise source:
\begin{align}
    \hat{o}_{10}^{(l)} &=  \frac{1}{\sqrt{2}} \big( \hat{j}_{10} + \hat{j}_{20} \big) - \tanh(r) \big( \hat{b}_0  -   \hat{a}_0^\dd \big), \\
    \hat{o}_{10}^{(r)} &=  \frac{1}{\sqrt{2}} \big( \hat{j}_{10} - \hat{j}_{20} \big) - \tanh(r)  \big( \hat{u}_0 - \hat{v}_0^\dd \big) .
\end{align}
In the regime where this noise is very large, the no-delay telemirror is thus mode-discriminating but not mode-selective; one mode arrives cleanly at Bob's output, while the other is polluted by noise. In Appendix \ref{appendixarbitrary}, we show how Alice can discriminate any arbitrary superposition mode (i.e.\ teleport it along with the entanglement resource) while polluting its orthogonal complement with noise. Meanwhile, the `$\perp$' output modes in the limit of high gain are given by $\hat{j}_{1\perp}' = - \hat{b}_{-\perp} $, $\hat{j}_{2\perp}' = \hat{b}_{+\perp} $; after mixing these together on another balanced beamsplitter, the original vacuum inputs $\hat{o}_{1\perp}^{(l)} = - \hat{e}_{1\perp}$ and $\hat{o}_{1\perp}^{(r)} = - \hat{u}_{1\perp}$ can be retrieved.  

For the reflected modes (after another tedious but straightforward calculation, shown in Appendix \ref{appendix1}), we obtain in the limit of high gain, 
\begin{align}
\begin{split} 
    \hat{r}_{10} &= \frac{1}{2\sqrt{2}} ( \hat{j}_{10}^\dd - \hat{j}_{20}^\dd ) - \hat{u}_0^\dd + \frac{3}{2}\hat{v}_0 , \\
    \hat{r}_{20} &= \frac{1}{2\sqrt{2}} ( \hat{j}_{10} - \hat{j}_{20} ) + \hat{u}_0 - \frac{1}{2} \hat{v}_0^\dd, \\
    \hat{r}_{30} &= \frac{1}{2\sqrt{2}} ( \hat{j}_{10}^\dd + \hat{j}_{20}^\dd ) + \frac{3}{2} \hat{a}_0 - \hat{b}_0^\dd, \\
    \hat{r}_{40} &= \frac{1}{2\sqrt{2}} ( \hat{j}_{10} + \hat{j}_{20} ) - \frac{1}{2} \hat{a}_0^\dd + \hat{b}_0 .
\end{split} 
\end{align}
The anti-symmetric superposition appears on the reflected side with additional noise from the vacuum inputs, while the symmetric superposition becomes buried beneath the noise from the entanglement resource modes with unbalanced coefficients. In the `$\perp$' modes, the leftover input vacuum modes appear at the reflected side as well:
\begin{align}
\begin{split} 
    \hat{r}_{1\perp} &= \hat{e}_{2\perp} , \vt \\
    \hat{r}_{2\perp} &= \hat{v}_{1\perp}, \vt 
\end{split}
\begin{split} 
    \hat{r}_{3\perp } &= \hat{j}_{1\perp }, \vt \\ 
    \hat{r}_{4\perp } &= \hat{j}_{2\perp } . \vt 
\end{split} 
\end{align}
The results from this section have affirmed those derived in Sec.\ \ref{timedelay}, namely that time-delays are a fundamental property of mode-selective mirrors. In particular, we attempted to circumvent this issue by transmitting (i.e.\ teleporting) each temporal component of the multi-mode input `on-the-run': that is, teleporting the temporal components independently, rather than waiting for the whole mode to enter the circuit. While this approach respects relativistic constraints (the propagation time of the mode is limited by the speed of light, and causality is preserved), it nevertheless fails as a completely mode-selective device. This property may be problematic in scenarios where one wishes to isolate a single-mode
for experimental purposes or in quantum causality problems \cite{Branciard_2015,abbottPhysRevA.94.032131}.

\section{Conclusion}\label{sec:VI}
The aims of this paper have been threefold. Firstly, we proposed a new teleportation model for the mode-selective mirror, which is widely used in quantum information and communication 
\cite{ansariPRXQuantum.2.010301,DonohuePhysRevLett.121.090501,WasilewskiPhysRevLett.99.123601}, relativistic QFT \cite{suPhysRevD.90.084022,onoePhysRevD.98.036011,onoePhysRevD.99.116001,FooPhysRevD.101.085006,Foo_2020, SuPhysRevX.9.011007} and experimental applications in quantum optics \cite{ansariPhysRevA.96.063817, Christ_2013,Brecht_2011,Eckstein_2011}. 

Next, we applied our new model to study in detail a well-known causality problem in the propagation of temporally extended modes through such mirrors. Specifically, we showed that mode-selective mirrors necessarily delay incoming modes based on the length of the mode itself. The underlying explanation of our result is closely connected with the pathological issues arising from instantaneous, nonlocal measurements in relativistic QFT \cite{Sorkin:1993gg}, which have been shown to elicit violations of causality when treated without care \cite{Lin:2013loa}. Finally, we investigated the issues which arise when one attempts to construct a mode-selective mirror with no time-delay. Intriguingly, such a mirror cannot be considered mode-selective, since it transmits the mode of choice \textit{and} an orthogonal superposition mode. Furthermore, an observer on the receiving side of the mirror cannot retrieve useful information about the temporal structure of the modes due to additional noise which propagates with them. 

Our novel approach to the mode-selective teleportation opens several pathways for future research. On the one hand it presents a new method of enacting mode-selective operations in quantum optics, where the prevailing techniques have been the quantum pulse gate and Raman quantum memory \cite{reimPhysRevLett.108.263602}. Achieving high efficiencies is a perpetual aim within the field of experimental quantum optics, and our approach provides a feasible path towards this end. The telemirror also links the quantum optics and quantum communication with the study of causality in relativistic quantum information \cite{Mann_2012} and quantum field theory. We have illuminated in a remarkably simple way the inherent issues underlying `impossible measurements' \cite{Sorkin:1993gg} in relativistic quantum field theory. While other works in this field have utilised theoretical techniques ranging from algebraic QFT \cite{fewster2020quantum} to measurement theory \cite{Lin:2013loa}, our approach frames these effects in quantum optical settings which have already been shown to be experimentally accessible.

\acknowledgments 
The authors thank Fabio Costa for fruitful discussions. This research was supported by the Australian Research Council Centre of Excellence for Quantum Computation and Communication Technology (Project
No. CE170100012) and DECRA Grant DE180101443.

\appendix

\section{Reflected mode for the telemirrors}\label{appendix1}
\subsection{Mode-selective telemirror}
\begin{figure}[h]
    \centering
    \includegraphics[width=0.425\linewidth]{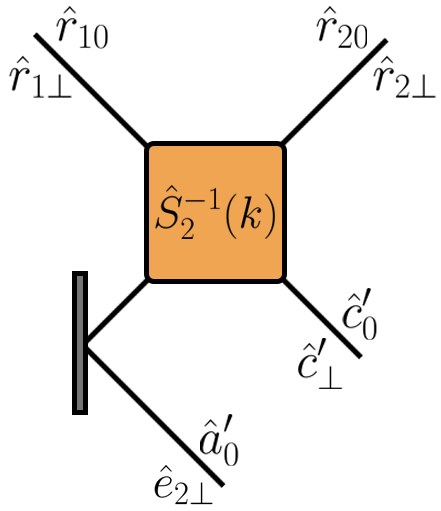}
    \caption{Circuit diagram for the operations enacted upon the reflected modes to retrieve the input vacua for the atemporal mode-selective telemirror. }
    \label{fig:reflectedoperations1}
\end{figure}

In this section, we derive the unitary transformation required to retrieve the input vacuum modes, $\hat{e}_{10}$ and $\hat{e}_{20}$, on the reflected side of the mode-selective (atemporal) telemirror from Sec.\ \ref{allopticaltelemirroratemporal}. As will be shown, these operations may be reduced to a single inverse squeezing operation, shown in Fig.\ \ref{fig:reflectedoperations1}. 
To derive the squeezing parameter $k$, it is instructive to decompose the single unitary operator $\hat{S}_2^{-1}(k)$ into a series of squeezers which reverse the initial operations performed on the input modes which generated (a) the entanglement resource modes, and (b) the classical channel. Recall that the output of the classical channel squeezer and the output of the beamsplitter displacement at Bob's output port are respectively given by 
\begin{align}
    \hat{a}_0' &= \cosh (r) \hat{a}_0 + \sinh (r) \hat{j}_0^\dd \vt \\
    \hat{c}_0' &= \text{sech}(r) \hat{b}_0 + \tanh(r) \cosh(r)\hat{j}_0 \non \vt  \\
    & + \tanh(r) \sinh(r) \hat{a}_0^\dd  . \vt  
\end{align}
We perform successive inverse squeezing operations on the output modes with respect to the parameters $r$ and $s$ used to generate the initial states:
\begin{align}
    \hat{q}_{10} &= \cosh (r) \hat{a}_0' - \sinh(r) \hat{c}_0'^\dd , \vt \\
    \hat{q}_{20} &= \cosh (r) \hat{c}_0' - \sinh(r) \hat{a}_0'^\dd , \vt 
\intertext{and then,}
    \hat{p}_{10} &= \cosh (s) \hat{q}_{10} - \sinh(s) \hat{q}_{20}^\dd, \vt \\
    \hat{p}_{20} &= \cosh (s) \hat{q}_{20} - \sinh(s) \hat{q}_{10}^\dd. \vt 
\end{align}
In the limit where $s$, $r\to \infty$, the reflected output modes reduce to 
\begin{align}\label{a7}
    \hat{p}_{10} &= \frac{5}{4} \hat{e}_{10} - \frac{3}{4} \hat{e}_{20}^\dd, \\     \label{a8} 
    \hat{p}_{20} &= \frac{5}{4} \hat{e}_{20} - \frac{3}{4} \hat{e}_{10}^\dd.
\end{align}
The $\hat{p}_{i0}$ modes form a two-mode squeezed state. To retrieve $\hat{e}_{10}$ and $\hat{e}_{20}$, we perform a final two-mode squeezing operation on $\hat{p}_{i0}$, with squeezing parameter $t = \text{arccosh}(5/4)$ which yields
\begin{align}
    \hat{r}_{10} &= \cosh(t) \hat{p}_{10} + \sinh(t) \hat{p}_{20}^{\prime\dd } = \hat{e}_{10} , \vt \\ 
    \hat{r}_{20} &= \cosh(t) \hat{p}_{20} + \sinh(t) \hat{p}_{10}^{ \dd } = \hat{e}_{20} . \vt 
\end{align}
This confirms that the teleportation protocol is fully unitary. The additional squeezing transformation is required due to the beamsplitter interaction at Bob's output port. Since some of the classical channel mode, $\hat{c}_0$, is contained in the teleported mode, $\hat{j}_0'$, retrieving $\hat{e}_{10}$ and $\hat{e}_{20}$ is more complicated than disentangling them via the successive inverse squeezing transformations. The three operations above can be combined into a single, inverse two-mode squeezing operation with squeezing parameter $k$, defined as 
\begin{align}
    k = \ln \Big( \cosh (t - r - s ) - \sinh (t - r - s ) \Big) 
\end{align}
so that
\begin{align}
    \hat{r}_{10} &= \cosh (k) \hat{a}_0' - \sinh(k) \hat{c}_0'^\dd = \hat{e}_{10}, \vt  \\
    \hat{r}_{20} &= \cosh (k) \hat{c}_0' - \sinh(k) \hat{a}_0'^\dd = \hat{e}_{20}, \vt 
\end{align}
as shown in Eq.\ (\ref{46}) in the main text. As already discussed in Sec.\ \ref{allopticaltelemirroratemporal}, the orthogonal mode $\hat{j}_\perp$ appears in the reflected mode $\hat{r}_{1\perp}$.  

\subsection{Temporal mode telemirror}
\begin{figure}[h]
    \centering
    \includegraphics[width=0.7\linewidth]{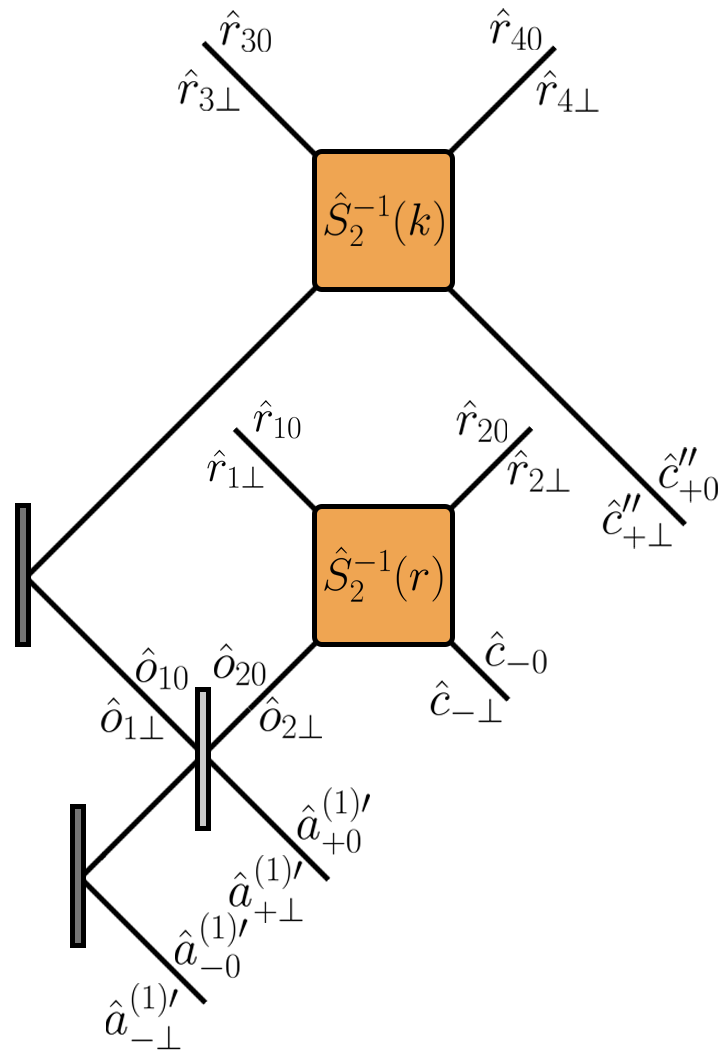}
    \caption{Circuit diagram for the operations used to retrieve the input vacua for the temporal mode time-delayed telemirror.}
    \label{fig:reflectedtemporalmode}
\end{figure}

We can perform a similar analysis to retrieve the reflected mode from the temporal mode telemirror. For simplicity, we study the scenario presented in Sec.\ \ref{allopticaltimedelay}, where the symmetric superposition of the input temporal modes was transmitted to Bob. Of course, this generalises to arbitrary mode-selection, shown in Appendix \ref{appendixarbitrarytemporalmode}. The operations enacted on the output modes are shown in Fig.\ \ref{fig:reflectedtemporalmode}.
Firstly, the output modes on the reflected side of the telemirror are given by 
\begin{align}
    \hat{a}_{-0} &= \frac{\cosh(r)}{\sqrt{2}} ( \hat{a}_0^\dd + ^\dd ) + \sinh(r) \hat{j}_{10}^\dd \vphantom{\sqrt{1 - \frac{1}{2\cosh^2(r)}}}, \\
    \hat{a}_{+0} &= \frac{\cosh(r)}{\sqrt{2}} ( \hat{a}_0 -  ) + \sinh(r) \hat{j}_{20}^\dd \vphantom{\sqrt{1 - \frac{1}{2\cosh^2(r)}}}, \\
    \hat{c}_{-0} &= \frac{\cosh(r)}{\sqrt{2}} (\hat{j}_{10} - \hat{j}_{20} ) + \sinh(r) \hat{v}_0^\dd \vphantom{\sqrt{1 - \frac{1}{2\cosh^2(r)}}}, \\
    \hat{c}_{+0}'' &= \left( \frac{\cosh(r)}{\sqrt{2}} - \frac{1}{2\sqrt{2}\cosh(r)} \right) ( \hat{j}_{10} + \hat{j}_{20} )  \non \\
    & + \frac{1}{2\cosh(r)} \left( \sqrt{1 - \frac{1}{2\cosh^2(r)}} - 1 \right)\hat{u}_0 \nonumber \\
    & +  \frac{1}{2\cosh(r)} \left( \sqrt{1 - \frac{1}{2\cosh^2(r)}} + 1 \right) \hat{b}_0 \non \\
    & + \left( \sinh(r) - \frac{\tanh(r)}{\cosh(r)} \right) \hat{a}_0^\dd .
\end{align} 
Mixing the $\hat{a}_{\pm0}$ modes at a balanced beamsplitter yields
\begin{align}
    \hat{o}_{10} &= \frac{1}{\sqrt{2}} ( \hat{a}_{+0}  + \hat{a}_{-0}  ) \non  \\
    &= \cosh(r) \hat{a}_0 + \frac{\sinh(r)}{\sqrt{2}} ( \hat{j}_{10}^\dd + \hat{j}_{20}^\dd ), \\
    \hat{o}_{20} &= \frac{1}{\sqrt{2}} ( \hat{a}_{-0}  - \hat{a}_{+0}  ) \non \\
    &= \cosh(r)\hat{v}_0   + \frac{\sinh(r)}{\sqrt{2}} ( \hat{j}_{10}^\dd - \hat{j}_{20}^\dd ) .
\end{align}
Mixing $\hat{c}_{-0}$ and $\hat{o}_{20}$ at an inverse squeezer with squeezing coefficient $r$ yields
\begin{align}
    \hat{r}_{10} &= \cosh(r) \hat{o}_{10} - \sinh(r) \hat{o}_{20}^\dd = \hat{v}_0  , \vt  \\
    \hat{r}_{20} &= \cosh(r) \hat{o}_{20} - \sinh(r) \hat{o}_{10}^\dd =  \frac{1}{\sqrt{2}} (\hat{j}_{10} - \hat{j}_{20} ) , \vt 
\end{align}
as shown in Eq.\ (\ref{74reflectedmode}). To retrieve the input vacua $\hat{e}_{i0}$ used to generate the EPR resource, one needs to apply the series of operations shown for the atemporal version of the protocol, which can be reduced to the action of a single unitary $\hat{S}_2(k)$. The operations are applied to the mode pair $\hat{o}_{10}$ and $\hat{c}_{+0}''$.

\section{Telefilters and telemirrors with imperfect efficiency}\label{appendix:imperfectelemirror}
\subsection{All-optical telemirror with finite squeezing}
\begin{figure}[h]
    \centering
    \includegraphics[width=0.9\linewidth]{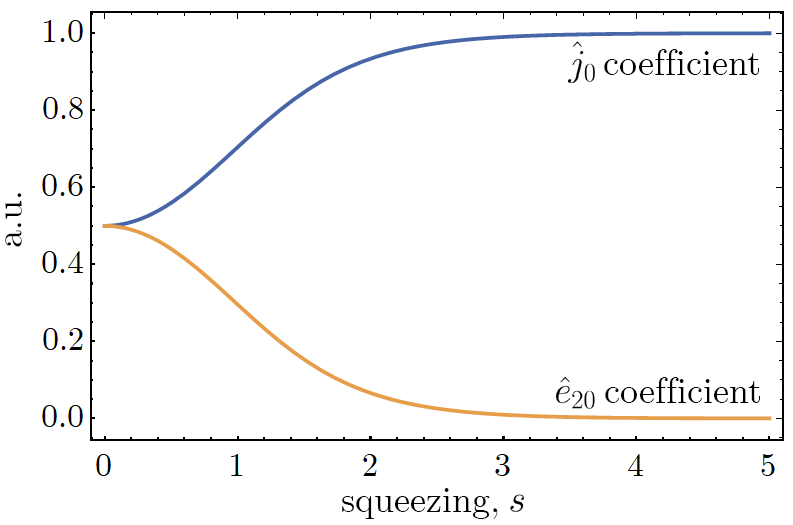}
    \caption{Plot of the coefficients of $\hat{j}_0$ and $\hat{e}_{20}$ in the output mode $\hat{j}_0'$}
    \label{fig:coefficients2}
\end{figure}

In Sec.\ \ref{sec:imperfectefficiency}, we presented a calculation for the atemporal mode-selective telefilter with finite squeezing. For completeness, we present here the input-output relations for the atemporal telemirror with finite-squeezing. The classical channel is generated via a two-mode squeezer which couples the input mode $\hat{j}_0$ with the entanglement resource mode, $\hat{a}_0$, with the gain of the channel controlled by the squeezing parameter $s$. That is, 
\begin{align}
    \hat{c}_0 &= \cosh(s) \hat{j}_0 + \sinh(s) \hat{a}_0^\dd  \vt \\
    \hat{a}_0' &= \cosh(s) \hat{a}_0 + \sinh(s) \hat{j}_0^\dd  . \vt 
\end{align}
Bob mixes $\hat{c}_0$ and $\hat{b}_0$ on a beamsplitter with transmission coefficient
\begin{align}
    \eta &= \frac{2\cosh^2(s)}{3 + \cosh(2s)}
\end{align}
so that
\begin{align}
    \hat{c}_0' &= \sqrt{\eta} \hat{b}_0 - \sqrt{1 -\eta} \hat{c}_0 \vphantom{\frac{\sqrt{2}}{\sqrt{3 + \cosh(2s)}}}, \\
    \hat{j}_0' &= \sqrt{\eta} \hat{c}_0 - \sqrt{1 -\eta} \hat{b}_0 \non  \\
    &= \frac{\sqrt{2}}{\sqrt{3 + \cosh(2s)}} ( \cosh(s) \hat{j}_0 - \hat{e}_{20} ) .
\end{align}
In Fig.\ \ref{fig:coefficients2}, we have plotted the modulus square of the mode coefficients in the output mode. We see clearly that for finite squeezing levels, the output mode transmit $\hat{j}_0$ with some imperfect efficiency. In the limit of perfect squeezing, one retrieves the result for the all-optical telemirror in Sec.\ \ref{allopticaltelemirroratemporal}. 
We can also calculate the reflected modes. To do this, we perform two successive inverse squeezing operations on the reflected modes $\hat{a}_0'$ and $\hat{c}_0'$ with squeezing parameter $s$, yielding
\begin{align}
    \hat{q}_{10} &= \cosh(s) \hat{a}_0' - \sinh(s) \hat{c}_0^{\prime\dd } \vt \\
    \hat{q}_{20} &= \cosh(s) \hat{c}_0' - \sinh(s) \hat{a}_0^{\prime \dd } \vt \\
    \hat{p}_{10} &= \cosh(s) \hat{q}_{10} - \sinh(s) \hat{q}_{20}^\dd  \vt \\
    \hat{p}_{20} &= \cosh(s) \hat{q}_{20} - \sinh(s) \hat{q}_{10}^\dd . \vt 
\end{align}
A final squeezing operation with the squeezing parameter set to $\text{arccosh}(5/4)$ retrieves the vacuum modes $\hat{e}_{i0}$ in the limit of infinite squeezing. For finite squeezing, $\hat{j}_0$ appears on one of the reflected modes, complementing the part of the mode which is transmitted; obviously, this vanishes as $s \to \infty$. 
% This is shown graphically in Fig.\ \ref{fig:coefficientmodes2}, where the $\hat{j}_0$ mode and its conjugate likewise vanish in the limit (i.e.\ as the transmitted mode approaches $\hat{j}_0$).
% \begin{figure}[h]
%     \centering
%     \includegraphics[width=0.9\linewidth]{FigApp2b coefficients telemirror finite squeezing.png}
%     \caption{Plot of the coefficients of the constituent modes in the reflected outputs, $\hat{p}_{10}$ (top) and $\hat{p}_{20}$ (bottom).}
%     \label{fig:coefficientmodes2}
% \end{figure}

\subsection{Time-delayed temporal mode telefilter with finite squeezing}\label{appendixtwomodefinite}
Here, we consider the temporal mode telefilter (with two input temporal modes) with finite squeezing. This can be straightforwardly generalised to an $N$-mode input. The equivalent calculation for the temporal mode telemirror is tedious but follows analogously. Firstly, we choose the effective gain of the classical channel to be $\zeta = \tanh(s)/(2 \sqrt{2}|\beta|)$ so that the total measurement operator is given by 
\begin{align}\label{a24}
    \hat{M} &= \frac{\tanh(s)}{2\sqrt{2}} ( \sqrt{2}  \hat{j}_{10} + \sqrt{2}\hat{j}_{20} + 2 \hat{a}_0^\dd ) .
\end{align} 
As before, we displace the output temporal components at Bob's side of the telemirror, yielding the results
\begin{align}
    \hat{j}_{10}' &= \frac{1}{2} \big( \hat{j}_{10} + \hat{j}_{20} \big) \tanh(s)  + \frac{1}{\sqrt{2}} \hat{u}_0 \VT \non \\
    & + \frac{\tanh(s) \sinh(s) - \cosh(s)}{\sqrt{2}} \hat{e}_{20} \VT , \\ 
    \hat{j}_{20}'&= - \frac{\tanh(s)}{2} \big( \hat{j}_{10} + \hat{j}_{20} \big) + \frac{1}{\sqrt{2}} \hat{u}_0 \non \VT \\
    & + \frac{\cosh(s) - \tanh(s) \sinh(s)}{\sqrt{2}} \hat{e}_{20} .
\end{align}
For clarity, it is instructive to mix $\hat{j}_{10}'$ and $\hat{j}_{20}'$ on a balanced beamsplitter, yielding
\begin{align}
    \hat{o}_{10}^{(l)} &= \frac{1}{\sqrt{2}} \big( \hat{j}_{10}'+ \hat{j}_{20}'\big) \VT \\
    &= \frac{ \tanh(s) }{\sqrt{2}} \big( \hat{j}_{10} + \hat{j}_{20} \big)- \frac{\hat{e}_{20}}{\cosh(s)} , \VT \\
    \hat{o}_{10}^{(r)} &= \frac{1}{\sqrt{2}} \big( \hat{j}_{10}'- \hat{j}_{20}'\big) =  \hat{u}_0 .
\end{align}
In this case, we see that the symmetric superposition of $\hat{j}_{10}$ and $\hat{j}_{20}$ has been teleported to Bob's output, with efficiency $\tanh(s)$ dictated by the amount of squeezing used to generate the entanglement-resource modes, $\hat{a}_0$ and $\hat{b}_0$. In the limit of perfect entanglement, we simply obtain the result shown in Eq.\ (\ref{42}) and (\ref{36}). To generalise this result to an $N$-mode input, one must appropriately choose the effective gain $\xi$ of the classical channel so that the channel between Alice and Bob is the identity -- see for example, the prefactor of Eq.\ (\ref{a24}). 

\section{Selecting an arbitrary superposition mode using the temporal mode telefilter and telemirror}\label{appendixarbitrary}
In the main text, we demonstrated simple examples where the telefilter and telemirror transmitted a particular superposition of the input temporal modes, and filtering or reflecting all others. Here, we generalise those result to show that any arbitrary mode can be selected by these devices, with the orthogonal modes filtered or reflected. We present the general calculations in the same order in which they appear in the main text (corresponding to Sec.\ \ref{sec:time-delaytemporalmodetelefilter}, \ref{allopticaltimedelay}, \ref{sec:nodelaytelefilter} and \ref{sec:nodelaytelemirror} respectively).

\subsection{Time-delayed temporal mode telefilter}\label{sec:C1}
To select an arbitrary mode with the time-delayed temporal mode telefilter, we require some additional degrees of freedom in our setup. Previously, the measurement operators were constructed from the $\hat{X}$ and $\hat{P}$ quadratures from the output modes at the dual homodyne measurement. In full generality, we can construct these measurement operators using quadrature operators at an arbitrary phase with a $\pi/2$ phase shift between them. That is,
\begin{align}
    \hat{X}_{a_{-0}} (\phi_1) &= \hat{a}_{-0}' e^{-i\phi_1} + \hat{a}_{-0}^{\prime\dd } e^{i\phi_1} \vt  \\
    \hat{X}_{j_{10'}}(\phi_1') &= \hat{j}_{10}' e^{-i\phi_1'} + \hat{j}_{10}^{\prime\dd } e^{i\phi_1'} \vt \\
    \hat{X}_{a_{+0}} (\phi_2) &= \hat{a}_{+0}e^{-i\phi_2} + \hat{a}_{+0}e^{i\phi_2} \vt \\
    \hat{X}_{j_{20}'}(\phi_2')  &= \hat{j}_{20}' e^{-i\phi_2'} + \hat{j}_{20}^{\prime\dd } e^{i\phi_2'}  \vt 
\end{align}
where $\phi_i' = \phi_i + \pi/2$. The measurement operators are now given by 
\begin{align}
    \hat{M}_1 &= \sqrt{2} |\beta| ( e^{-i\phi_1 } \hat{j}_{10} - i \sqrt{1 - \alpha} e^{i(\phi + \phi_1)} \hat{a}_0^\dd - \sqrt{\alpha} e^{i\phi_1} \hat{v}_0^\dd ) , \vt  \\
    \hat{M}_2 &= \sqrt{2} |\beta| ( e^{-i\phi_2 } \hat{j}_{20} - \sqrt{\alpha} e^{i\phi_2} \hat{a}_0^\dd - i \sqrt{1- \alpha} e^{-i(\phi - \phi_2) } \hat{v}_0^\dd  ). \vt 
\end{align}
We add these together in the usual way, 
\begin{align}
    \hat{M} &= \zeta_1 (\phi_1,\alpha) \hat{M}_1 + \zeta_2(\phi_2, \alpha) \hat{M}_2 
\end{align}
with the following choices for the weighting of the individual measurements:
\begin{align}
    \zeta_1(\phi_1, \alpha) &= \frac{1 - \alpha}{\sqrt{2}\lambda_1} e^{-i\phi_1 } \\
    \zeta_2(\phi_2, \alpha) &= \frac{\sqrt{\alpha(1 - \alpha)}}{\sqrt{2}\lambda_1} e^{-i\phi_2} .
\end{align}
We displace Bob's entanglement resource modes using the total measurement operator
\begin{align}
    \hat{j}_{10}' &= \hat{b}_{-0} + \lambda_1 (\alpha) \hat{M} \vt \\
    \hat{j}_{20}' &= \hat{b}_{+0} + \lambda_2(\alpha) \hat{M}  \vt 
\end{align}
with the relationship
\begin{align}
    \lambda_2(\alpha) &= \frac{\lambda_1 \sqrt{\alpha}}{\sqrt{1-\alpha}} .
\end{align}
Mixing the output modes on a passive beamsplitter yields
\begin{align}
    \hat{o}_{10}^{(l)} &= \sqrt{\alpha} \hat{j}_{20}' + \sqrt{1 - \alpha} \hat{j}_{10}'  \non 
    \vt  \\ 
    &= e^{-2i\phi_1}\sqrt{1 - \alpha} \hat{j}_{10} + e^{-2i\phi_2 } \sqrt{\alpha} \hat{j}_{20} \non \vt \\
    & + ( \hat{b}_0 - \hat{a}_0^\dd )  \vt \\
    \hat{o}_{10}^{(r)} &= \sqrt{\alpha} \hat{j}_{10}'  - \sqrt{1 - \alpha} \hat{j}_{20}' \non \vt  \\
    &= \hat{u}_0 \vphantom{ e^{-2i\phi_1}\sqrt{1 - \alpha} \hat{j}_{10} + e^{-2i\phi_2 } \sqrt{\alpha} \hat{j}_{20} + ( \hat{b}_0 - \hat{a}_0^\dd )}. \vt 
\end{align}
By tuning the quadrature phases $\phi_i$ and the beamsplitter transmission coefficient $\alpha$, Alice can select any arbitrary mode to transmit to Bob in $\hat{o}_{10}^{(l)}$, while the other mode contains vacuum.

\subsection{Time-delayed temporal mode telemirror}\label{appendixarbitrarytemporalmode}
An analogous calculation can be performed for the time-delayed temporal mode telemirror. The extra degrees of freedom we introduce are phase shifts on the classical channel modes, namely
\begin{align}
    \hat{U}^\dd (\phi_j) \hat{j} \hat{U}(\phi_j) = \hat{j} e^{i\phi_j} 
\end{align}
where $\hat{j} = \hat{c}_{i0}$, $\hat{a}_{\pm 0}'$. To obtain the desired output, let us impose the following constraints on the beamsplitter phases:
\begin{align}
\begin{split} 
    \theta_- &= \phi + \theta_+ + \frac{\pi}{2}, \\
    \theta_+ &= \phi_{c_{20}} + \frac{\pi}{2}, 
\end{split} 
\begin{split} 
    \phi_{c_{10}} &= - \phi + \frac{\pi}{2}, \\
    \phi_{c_0} &= \phi + \phi_{c_{10}} - \phi_{c_{20}} \vphantom{\frac{\pi}{2}}.
\end{split} 
\end{align}
Another set of constraints could have been chosen, since the choice above is not unique. This yields the output modes
\begin{align}
    \hat{j}_{10}' &= (1 - \alpha)e^{-2i\phi } \hat{j}_{10} + i e^{-i\phi } \sqrt{\alpha ( 1- \alpha)} \hat{j}_{20} \non \vt  \\
    & - i e^{-i\phi } \sqrt{1 - \alpha} ( \hat{b}_0 - \hat{a}_0^\dd )  + \sqrt{\alpha} \hat{u}_0 \vt \\
    \hat{j}_{20}' &= i e^{-i\phi } \sqrt{\alpha( 1- \alpha ) } \hat{j}_{10} - \alpha \hat{j}_{20} + \sqrt{\alpha} ( \hat{b}_0 - \hat{a}_0^\dd ) \non \vt \\
    & - i e^{i\phi} \sqrt{1 - \alpha} \hat{u}_0 . \vt 
\end{align}
These output modes are mixed at a final passive beamsplitter, 
\begin{align}
    \hat{o}_{10}^{(l)} &= \sqrt{\alpha} \hat{j}_{20}' - i e^{-i\chi } \sqrt{1 - \alpha} \hat{j}_{10}'  \vt \\
    \hat{o}_{10}^{(r)} &= \sqrt{\alpha} \hat{j}_{10}'  - i e^{i\chi } \sqrt{1- \alpha} \hat{j}_{20}' . \vt 
\end{align}
By taking $\chi = - \phi + \pi$ we obtain the desired result, 
\begin{align}
    \hat{o}_{10}^{(l)} &= i e^{-i\phi } \sqrt{1 - \alpha} \hat{j}_{10} - \sqrt{\alpha} \hat{j}_{20} + ( \hat{b}_0 - \hat{a}_0^\dd ) , \vt \\
    \hat{o}_{10}^{(r)} &= \hat{u}_0 \vt  .
\end{align}
Thus, Alice may select any arbitrary mode to be transmitted to Bob. For the reflected modes, we firstly mix the outputs from the two-mode squeezers at a passive beamsplitter:
\begin{align}
    \hat{o}_{10} &= \sqrt{\mu} \hat{a}_{-0}^{\prime } - i e^{-i\phi_{a_{+0}}} \sqrt{1 - \mu } \hat{a}_{+0}^{\prime } \vt \\
    \hat{o}_{20} &= \sqrt{\mu} \hat{a}_{+0}^{\prime } - i e^{i\phi_{a_{+0}}} \sqrt{1 - \mu} \hat{a}_{-0}^{\prime } . \vt 
\end{align}
We then inverse squeeze $\hat{o}_{20}$ with the output from the mixing of the classical channel modes, $\hat{c}_{-0}$:
\begin{align}
    \hat{r}_{10} &= \cosh (r) \hat{o}_{20} - \sinh(r) \hat{c}_{-0}^\dd \vt, \\
    \hat{r}_{20} &= \cosh(r) \hat{c}_{-0} - \sinh(r) \hat{o}_{20}^\dd . \vt 
\end{align}
Like before, we can make the following choices for the beamsplitter phases, 
\begin{align}
    \phi_{c_{10}} &= - \phi + \frac{\pi}{2} , \qquad 
    \phi_{a_{+0}} = \phi - \phi_{a_{-0}} \vphantom{\frac{\pi}{2}} . 
\end{align}
This yields the output modes, 
\begin{align}
    \hat{r}_{10} &= - i e^{i\phi } \hat{v}_{0} \vt, \\
    \hat{r}_{20} &= i e^{-i\phi } \sqrt{\alpha} \hat{j}_{10} + \sqrt{1 - \alpha} \hat{j}_{20}.  \vt
\end{align}
The mode $\hat{r}_{20}$ is orthogonal to $\hat{o}_{10}^{(l)}$, giving the desired mode-selective property of the result.

\subsection{No-delay temporal mode telefilter}
For completeness, we show how the no-delay telefilter and telemirror can be generalised to partially select an arbitrary mode of interest. As before, we define our quadrature operators with the generic phase $\phi_i$. We construct the measurement operators
\begin{align}
    \hat{M}_1 &= \hat{X}_{a_{-0}}(\phi_1) + i \hat{X}_{j_{10}'} (\phi_1'') , \vt \\
    \hat{M}_2 &= \hat{X}_{a_{+0}}(\phi_2) + i \hat{X}_{j_{20}'}(\phi_2'' ) , \vt 
\end{align}
which are used to displace the entanglement resource modes, 
\begin{align}
    \hat{j}_{10}' &= \hat{b}_{-0} + \lambda_1(\phi_1) \hat{M}_1 , \vt \\
    \hat{j}_{20}' &= \hat{b}_{+0} + \lambda_2 ( \phi_2 ) \hat{M}_2 . \vt 
\end{align}
We obtain the output modes, 
\begin{align}
    \hat{j}_{10}' &= e^{-2i\phi_1 } \hat{j}_{10} + \sqrt{1 - \alpha}( \hat{b}_0 - \hat{a}_0^\dd ) \non \vt \\
    & + \sqrt{\alpha} (\hat{u}_0 - \hat{v}_0^\dd ), \vt \\
    \hat{j}_{20}' &= e^{-2i\phi_2 } \hat{j}_{20} + \sqrt{\alpha} (\hat{b}_0 - \hat{a}_0^\dd ) ,  \vt 
\end{align}
where we have retained the nomenclature for the quadrature phases in Sec.\ \ref{sec:C1}. Mixing these back on a passive beamsplitter in the usual way yields
\begin{align}
    \hat{o}_{10}^{(l)} &= e^{-2i\phi_1 } \sqrt{1 - \alpha} \hat{j}_{10} + e^{-2i\phi_2 } \sqrt{\alpha} \hat{j}_{20} \non \vt \\ 
    & + ( \hat{b}_0 - \hat{a}_0^\dd ) \vt \\
    \hat{o}_{10}^{(r)} &= e^{-2i\phi_1 } \sqrt{\alpha} \hat{j}_{10} - e^{-2i\phi_2 } \sqrt{1 - \alpha} \hat{j}_{20} \non \vt \\
    & + (\hat{u}_0 - \hat{v}_0^\dd )  \vt . 
\end{align}
Now the mode of interest with phases $\phi_i$ and transmission coefficient $\alpha$ appears in $\hat{o}_{10}^{(l)}$ with its complement in the other transmitted mode $\hat{o}_{10}^{(r)}$.

\subsection{No-delay temporal mode telemirror}
For the no-delay temporal mode telemirror, Alice could generate the squeezing classical channels with squeezing phases $\theta_{s_i}$
\begin{align}
    \hat{c}_{10} &= \cosh (r) \hat{j}_{10} - e^{i\phi{c_{10}}} \sinh(r) \hat{a}_{-0}^\dd  \vt , \\
    \hat{c}_{20} &= \cosh(r) \hat{j}_{20} - e^{i\phi{c_{20}}} \sinh(r) \hat{a}_{+0}^\dd . 
\end{align}
We perform the usual displacement of Bob's individual entanglement resource modes using these channel modes:
\begin{align}
    \hat{j}_{10}' &= \sqrt{\eta_-} \hat{b}_{-0} - ie^{i\theta_-} \sqrt{1 - \eta_-} \hat{c}_{10} \vt \\
    \hat{j}_{20}' &= \sqrt{\eta_+} \hat{b}_{+0} - i e^{i\theta_+} \sqrt{1 - \eta_+}\hat{c}_{20}. \vt 
\end{align}
As before, we can make simplifications by choosing the beamsplitter phases, knowing the desired outcome of the partially mode-selective mirror. For example, taking
\begin{align} 
\phi &= - \pi/2, \vt \\ 
\phi_{c_{10}} &= \pi /2 - \theta_- \vt 
\\
\phi_{c_{20}} &= \theta_- - \theta_+ + \phi_{c_{10}} , \vt
\end{align}
yields the outputs (in the limit of high gain, $r \to \infty$),
\begin{align}
    \hat{j}_{10}' &= - ie^{i\theta_- } \hat{j}_{10} + \sqrt{1- \alpha} ( \hat{b}_0 - \hat{a}_0^\dd ) \non \vt \\
    & + \sqrt{\alpha} ( \hat{u}_{10} - \hat{v}_{10}^\dd )  \vt \\
    \hat{j}_{20}' &= - ie^{i\theta_+ } \hat{j}_{20} + \sqrt{\alpha} ( \hat{b}_0 - \hat{a}_0^\dd ) \non \vt \\
    & - \sqrt{1- \alpha} (\hat{u}_{10} - \hat{v}_{10}^\dd )  . \vt 
\end{align}
Mixing these on a final beamsplitter yields
\begin{align}
    \hat{o}_{10}^{(l)} &= - i e^{i\theta_- } \sqrt{1 - \alpha} \hat{j}_{10} - i e^{i\theta_+ } \sqrt{\alpha} \hat{j}_{20} \non \vt \\
    & + ( \hat{b}_0 - \hat{a}_0^\dd ) \vt  \\
    \hat{o}_{10}^{(r)} &= - ie^{i\theta_- } \sqrt{\alpha} \hat{j}_{10} + i e^{i\theta_+} \sqrt{1 - \alpha} \hat{j}_{20} \non \vt 
    \\
    & + ( \hat{u}_{10} - \hat{v}_{10}^\dd )  \vt . 
\end{align}
Thus, Alice can select an arbitrary mode to be mode-discriminated, with the orthogonal superposition transmitted along with noise. We are less interested in the reflected modes in this case, since device doesn't isolate the mode of interest on either the transmitted or reflected sides. 

\section{$N$-mode generalisation}\label{sec:nmode}
We now generalise the temporal mode telefilter protocols studied in the main text to an input of $N$ temporal modes. In the large-$N$ limit, the input approximates an input wavepacket decomposed into an large number of orthogonal modes. Although the telefilter and telemirror protocols possess a direct mapping between one another, the calculations for the telefilter are much simpler, particularly for the time-delayed case. For brevity, we only present the $N$-mode extension of the telefilter, in both the time-delayed and no-delay cases.

\subsection{$N$-mode temporal mode time-delayed telefilter}
\begin{widetext}
We have the circuit shown in Fig.\ \ref{fig:nmodedelay}. This is the $N$-mode extension of the temporal mode time-delayed telefilter discussed in Sec.\ \ref{sec:time-delaytemporalmodetelefilter}. The $N$ temporal mode inputs are mixed with $N$ respective local oscillators, which are likewise mode-matched with the entanglement resource modes, $\hat{a}_{-0}^{(n)}\hdots \hat{a}_{+0}^{(N-1)}$. 
\begin{figure}[h]
    \centering
    \includegraphics[width=0.6\linewidth]{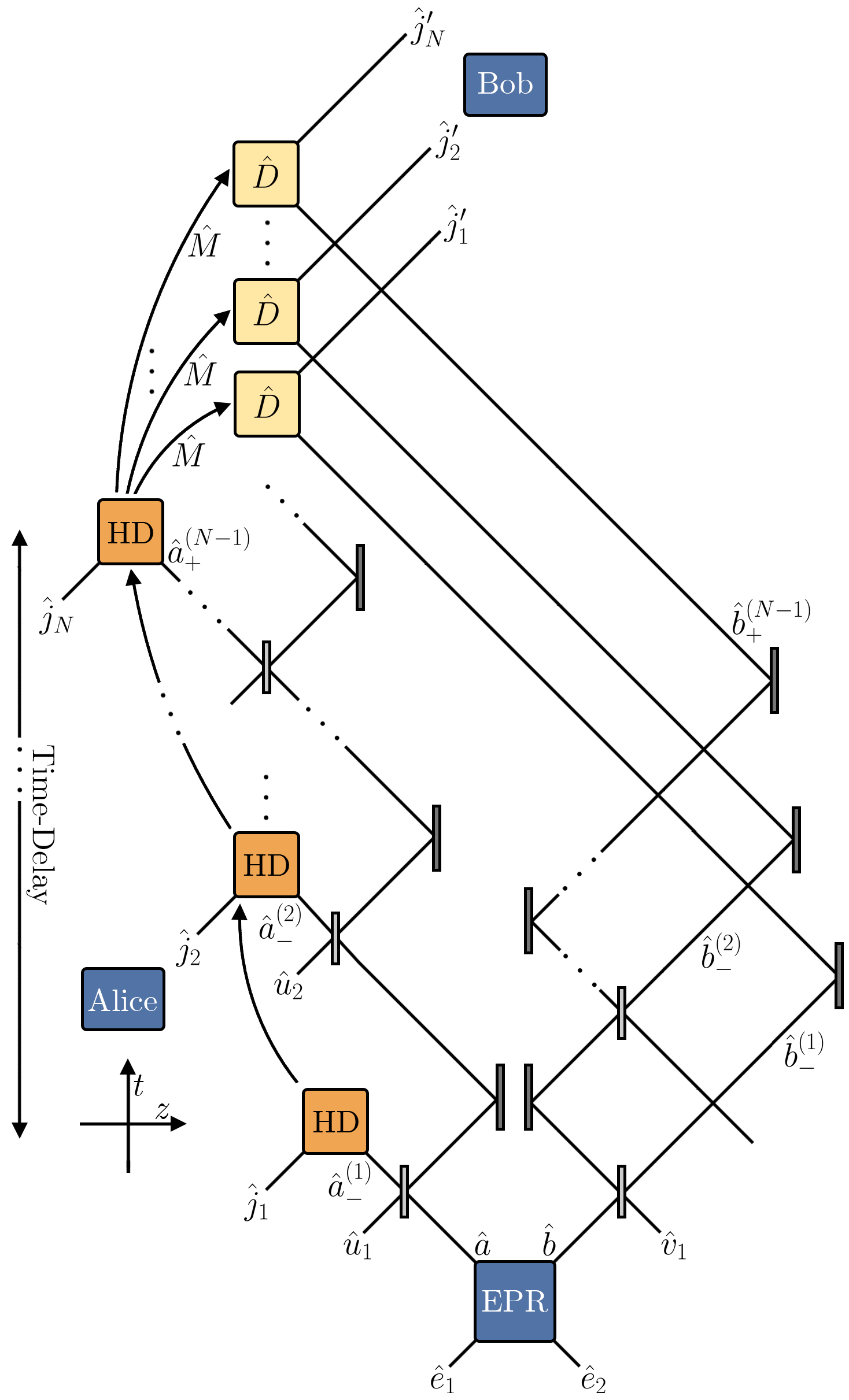}
    \caption{Circuit diagram for the $N$-mode temporal mode time-delayed telefilter. }
    \label{fig:nmodedelay}
\end{figure}
These entanglement resource modes are distributed among the $N$ temporal modes via passive beamsplitters with transmission coefficient $\alpha_i$ and phase $\phi_i$ as follows:
\begin{align}
\begin{split}
    \hat{a}_{-0}^{(1)} &= \sqrt{\alpha_1} \hat{v}_{10} - i e^{-i\phi_1 } \hat{a}_0 \\
    \hat{a}_{+0}^{(1)} &= \sqrt{\alpha_1}\hat{a}_0 - i e^{i\phi_1 } \sqrt{1 - \alpha_1}\hat{v}_{10} \vt \\
    \hat{a}_{-0}^{(2)} &= \sqrt{\alpha_2} \hat{v}_{20} - i e^{-i\phi_2 } \hat{a}_{+0}^{(1)} \vt \\
    \hat{a}_{+0}^{(2)} &= \sqrt{\alpha_2} \hat{a}_{+0}^{(1)} - i e^{i\phi_2 } \sqrt{1 - \alpha_2} \hat{v}_{20} \vt  \\
    &\:\:\: \vdots \non \\
    \hat{a}_{-0}^{(n)} &= \sqrt{\alpha_n} \hat{v}_{n0} - i e^{-i\phi_n } \sqrt{1 - \alpha_n} \hat{a}_{+0}^{(n-1) }  \vt \\
    \hat{a}_{+0}^{(n)} &= \sqrt{\alpha_n} \hat{a}_{+0}^{(n-1)} - i e^{i\phi_n } \sqrt{1 -\alpha_n} \hat{v}_{n0} \vt \\
    & \:\:\: \vdots \non \\
    \hat{a}_{-0}^{(N-1)} &= \sqrt{\alpha_{N-1}} \hat{v}_{(N-1)0} - i e^{-i\phi_{N-1} } \sqrt{1 - \alpha_{N-1}} \hat{a}_{+0}^{(N-2)} \vt \\
    \hat{a}_{+0}^{(N-1)} &= \sqrt{\alpha_{N-1}} \hat{a}_{+0}^{(N-2)} - i e^{i\phi_{N-1} } \sqrt{1 -\alpha_{N-1}} \hat{v}_{(N-1)0}  \vt 
\end{split}
\begin{split}
    \hat{b}_{-0}^{(1)} &= \sqrt{\alpha_1} \hat{u}_{10} - i e^{-i\phi_1 } \hat{b}_0 \vt \\
    \hat{b}_{+0}^{(1)} &= \sqrt{\alpha_1}\hat{b}_0 - i e^{i\phi_1 } \sqrt{1 - \alpha_1}\hat{u}_0 \vt \\
    \hat{b}_{-0}^{(2)} &= \sqrt{\alpha_2} \hat{u}_{20} - i e^{-i\phi_2 } \hat{b}_{+0}^{(1)} \vt \\
    \hat{b}_{+0}^{(2)} &= \sqrt{\alpha_2} \hat{b}_{+0}^{(1)} - i e^{i\phi_2 } \sqrt{1 - \alpha_2} \hat{u}_{20} \vt \\
    &\:\:\: \vdots \non \\
    \hat{b}_{-0}^{(n)} &= \sqrt{\alpha_n} \hat{u}_{n0} - i e^{-i\phi_n } \sqrt{1 - \alpha_n} \hat{b}_{+0}^{(n-1) } \vt  \\
    \hat{b}_{+0}^{(n)} &= \sqrt{\alpha_n} \hat{b}_{+0}^{(n-1)} - i e^{i\phi_n } \sqrt{1 -\alpha_n} \hat{u}_{n0} \vt \\
    & \:\:\: \vdots \non \\
    \hat{b}_{-0}^{(N-1)} &= \sqrt{\alpha_{N-1}} \hat{u}_{(N-1)0} - i e^{-i\phi_{N-1} } \sqrt{1 - \alpha_{N-1}} \hat{b}_{+0}^{(N-2)} \vt \\
    \hat{b}_{+0}^{(N-1)} &= \sqrt{\alpha_{N-1}} \hat{b}_{+0}^{(N-2)} - i e^{i\phi_{N-1} } \sqrt{1 -\alpha_{N-1}} \hat{u}_{(N-1)0}  . \vt
\end{split}
\end{align}
Some fraction of the original entanglement, $\hat{a}_0$ and $\hat{b}_0$, is distributed to each temporal mode and used to obtain the measurement result. At this point, let us specialise to $\phi_i = - \pi/2$ for the beamsplitter phases for simplicity. Of course, one can choose this arbitrarily to select a different mode -- this then requires a specifically tuned recombination of the output modes on Bob's transmitted side. Here, we simply present an example which illustrates the generic process for selecting a mode to-be-transmitted from Alice to Bob. For further simplification, we will take the measurement results to be 
\begin{align}
    \hat{M}_1 &= |\beta| (\hat{X}_{a_{-0}^{(1)}} + i \hat{P}_{j_{20}'} ) = \sqrt{2}|\beta| ( \hat{j}_{10} - \sqrt{1 - \alpha}\hat{a}_0^\dd - \sqrt{\alpha_1} \hat{v}_{10}^\dd ) \vt  \\
    \hat{M}_2 &= |\beta|( \hat{X}_{a_{-0}^{(2)\prime}} + i \hat{P}_{j_{20}'}) = \sqrt{2}|\beta| ( \hat{j}_{20} - \sqrt{\alpha(1 - \alpha_2)} \hat{a}_0^\dd + \sqrt{(1 - \alpha_1)( 1- \alpha_2) } \hat{v}_{10}^\dd - \sqrt{\alpha_2} \hat{v}_{20}^\dd ) \vt \\
    & \:\:\: \vdots \non \\
    \hat{M}_n &= |\beta|( \hat{X}_{a_{-0}^{(n)\prime}} + i \hat{P}_{j_{(n)0}'} ) = \sqrt{2} |\beta| ( \hat{j}_{n0} - \sqrt{\alpha_1 \alpha_2 \hdots (1 - \alpha_n)} \hat{a}_0^\dd + \sqrt{(1 - \alpha_1) \alpha_2 \hdots \alpha_{n-1} ( 1- \alpha_n)} \hat{v}_{10}^\dd \non \vt \\
    & + \sqrt{(1 - \alpha_2 ) \alpha_3 \hdots \alpha_{n-1} ( 1- \alpha_n)} \hat{v}_{20}^\dd + \hdots + \sqrt{(1 - \alpha_{n-1}) ( 1- \alpha_n) } \hat{v}_{(n-1)0}^\dd- \sqrt{\alpha_n}  \hat{v}_{(n)0}^\dd ) \vt  \\
    & \:\:\: \vdots \non \\ 
    \hat{M}_N &= |\beta| ( \hat{X}_{a_{+0}^{(N-1)\prime}} + i \hat{P}_{j_{(N)0}'} ) = \sqrt{2} |\beta| ( \hat{j}_{(N)0} - \sqrt{\alpha_1 \hdots \alpha_{N-1}} \hat{a}_0^\dd + \sqrt{(1 - \alpha_1) \alpha_2 \hdots \alpha_{N-1}} \hat{v}_{10}^\dd \non \vt \\
    & + \sqrt{(1 - \alpha_2 ) \alpha_3 \hdots \alpha_{N-1} } \hat{v}_{20}^\dd  + \hdots + \sqrt{1 - \alpha_{N-1}} \hat{v}_{(N-1)0}^\dd ) \vt .
\end{align}
We add these to form a total measurement operator, $\hat{M} = \sum_{i=1}^N \zeta_i \hat{M}_i$, where 
\begin{align}
    \zeta_1 &= \frac{\zeta_N \sqrt{1 - \alpha_1}}{ \sqrt{\alpha_1 \alpha_2 \hdots \alpha_{N-1}}} , \:\:\: 
    \zeta_2 = \frac{\zeta_N\sqrt{1- \alpha_2}}{\sqrt{\alpha_2 \alpha_3 \hdots \alpha_{N-1}}} , \:\:\: \hdots \:\:\: \zeta_n = \frac{\zeta_N\sqrt{1 -\alpha_n}}{\sqrt{\alpha_n \alpha_{n+1} \hdots \alpha_{N-1}}}, \:\:\: \hdots \:\:\:  
    \zeta_{N-1} = \frac{\zeta_N\sqrt{1 - \alpha_{N-1}}}{\sqrt{\alpha_{N-1}}}.
\end{align}
As shown in Fig.\ \ref{fig:nmodedelay}, one may partition the input field into a larger number of time-bin modes to obtain a better approximation to a continuous wavepacket. Overall, the total time-delay is the difference in the arrival times of the first and last time-bin modes. Now, the total measurement operator is used to displace the $N$ entanglement resource modes sent to Bob's side of the telefilter, 
\begin{align}
    \hat{j}_{10}' &= \hat{b}_{-0}^{(1)} + \lambda_1 \hat{M} ,\:\:\: 
    \hat{j}_{20}' = \hat{b}_{-0}^{(2)} + \lambda_2 \hat{M}_2 \:\:\: \hdots \:\:\: \hat{j}_{(n)0} = \hat{b}_{-0}^{(n)} + \lambda_n \hat{M}_n \:\:\: \hdots \:\:\: \hat{j}_{(N)0}' = \hat{b}_{+0}^{(N-1)} + \lambda_N \hat{M}_N \VT 
\end{align}
where we need to choose the effective gains in a likewise weighted fashion,
\begin{align}
    \lambda_1 &= \frac{\sqrt{(1 - \alpha_1) \alpha_1 \alpha_2 \hdots \alpha_{N-1}}}{|\beta| \sqrt{2}\zeta_N} , \:\:\: 
    \lambda_2 = \frac{\alpha_1 \sqrt{(1 - \alpha_2 ) \alpha_2 \hdots \alpha_{N-1}}}{|\beta|\sqrt{2}\zeta_N}, \:\:\: \hdots \:\:\:
    \lambda_n = \frac{\alpha_1 \alpha_2 \hdots \alpha_{n-1} \sqrt{(1 - \alpha_n) \alpha_n \hdots \alpha_{N-1}}}{|\beta| \sqrt{2} \zeta_N}, \\
    \lambda_N &= \frac{\alpha_1 \alpha_2 \hdots \alpha_{N-1} \alpha_N}{|\beta| \sqrt{2} \zeta_N}.
\end{align}
\begin{figure}[h]
    \centering
    \includegraphics[width=0.6\linewidth]{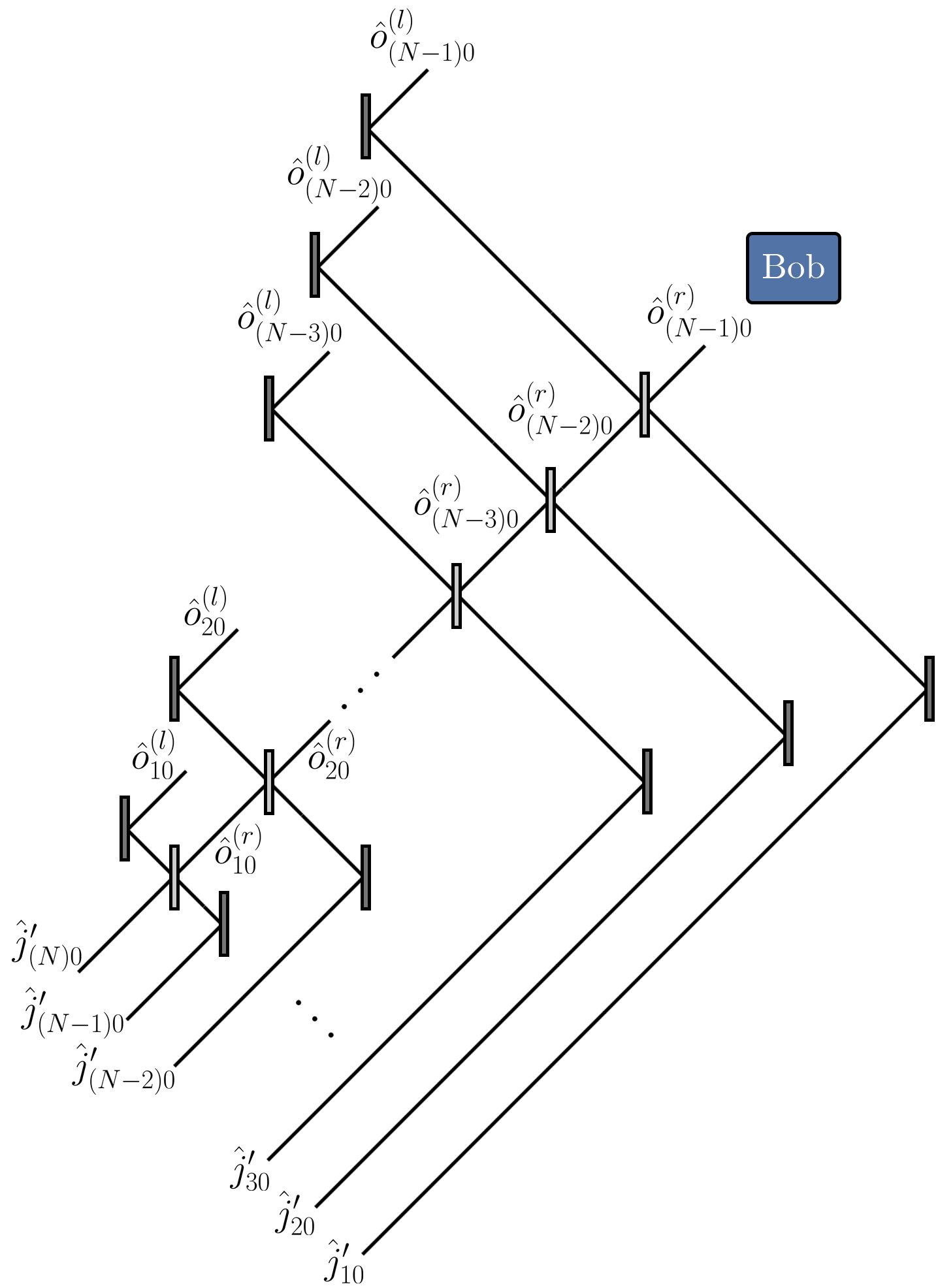}
    \caption{Circuit for the recombination of the output modes used to retrieve the mode of interest. }
    \label{fig:recombine}
\end{figure}
To retrieve the teleported mode, we mix these back together on weighted beamsplitters (see Fig.\ \ref{fig:recombine})  as follows:
\begin{align}
    \begin{split}
        \hat{o}_{10}^{(l)} &= \sqrt{\alpha_{N-1}} \hat{j}_{(N-1)0}' - \sqrt{1 - \alpha_{N-1}} \hat{j}_{(N)0}' \vt \\
        \hat{o}_{20}^{(l)} &= \sqrt{\alpha_{N-2}} \hat{j}_{(N-2)0}' - \sqrt{1 - \alpha_{N-2}}\hat{o}_{10}^{(r)} \vt \\
        & \:\:\: \vdots \non  \\
        \hat{o}_{(n)0}^{(l)} &= \sqrt{\alpha_{N-n}} \hat{j}_{(N-n)0}' - \sqrt{1 - \alpha_{N-n}} \hat{o}_{(n-1)0}^{(r)} \vt \\
        & \:\:\: \vdots \non \\
        \hat{o}_{(N-1)0}^{(l)} &= \sqrt{\alpha_1} \hat{j}_{10}' - \sqrt{1 - \alpha_1} \hat{o}_{(N-1)0}^{(r)} \vt 
    \end{split}
    \begin{split}
        \hat{o}_{10}^{(r)} &= \sqrt{\alpha_{N-1}} \hat{j}_{(N)0}' + \sqrt{1- \alpha_{N-1} } \hat{j}_{(N-1)0}' \\
        \hat{o}_{20}^{(r)}  &= \sqrt{\alpha_{N-2}}\hat{o}_{10}^{(r)} + \sqrt{1 - \alpha_{N-2}} \hat{j}_{(N-2)0}' \\
        & \:\:\: \vdots \non \\
        \hat{o}_{(n)0}^{(r)} &= \sqrt{\alpha_{N-n}} \hat{o}_{(n-1)0}^{(r)} + \sqrt{1 - \alpha_{N-n}} \hat{j}_{(N-n)0}' \\
        & \:\:\: \vdots \non \\
        \hat{o}_{(N-1)0}^{(r)} &= \sqrt{\alpha_1} \hat{o}_{(N-1)0}^{(r)} + \sqrt{1 - \alpha_1} \hat{j}_{10}'
    \end{split}
\end{align}
which yields the outputs
\begin{align*}
    \hat{o}_{10}^{(l)} &= \hat{u}_{(N-1)0} \\
    \hat{o}_{20}^{(l)} &= \hat{u}_{(N-2)0} \\
    & \:\:\: \vdots \non \\
    \hat{o}_{(N-2)0}^{(l)} &= \hat{u}_{20} \\ 
    \hat{o}_{(N-1)0}^{(l)} &=  \sqrt{1 -\alpha } \hat{j}_{10} + \sqrt{\alpha(1 - \alpha)} \hat{j}_{20} + \sqrt{\alpha_1 \alpha_2( 1- \alpha_3) } \hat{j}_{30} \non \\
    & + \hdots + \sqrt{\alpha_1 \hdots \alpha_{N-2} ( 1- \alpha_{N-1} )} \hat{j}_{(N-1)0} + \sqrt{\alpha_1 \hdots \alpha_{N-1}} \hat{j}_{(N)0} \\
    \hat{o}_{(N-1)0}^{(r)} &= \hat{u}_{10} 
\end{align*}
All of the output modes except $\hat{o}_{(N-1)0}^{(l)}$ contain the vacuum modes $\hat{u}_{(n)0}$. Only $\hat{o}_{(N-1)0}^{(l)}$ contains the selected superposition of the $N$ temporal modes (here, with equal phases between the constituent modes), which is the desired action of the mode-selective telefilter.

\subsection{$N$-mode temporal mode no-delay telefilter}
We have the circuit shown in Fig.\ \ref{fig:nmodenodelay}. 
\begin{figure}[h]
    \centering
    \includegraphics[width=0.5\linewidth]{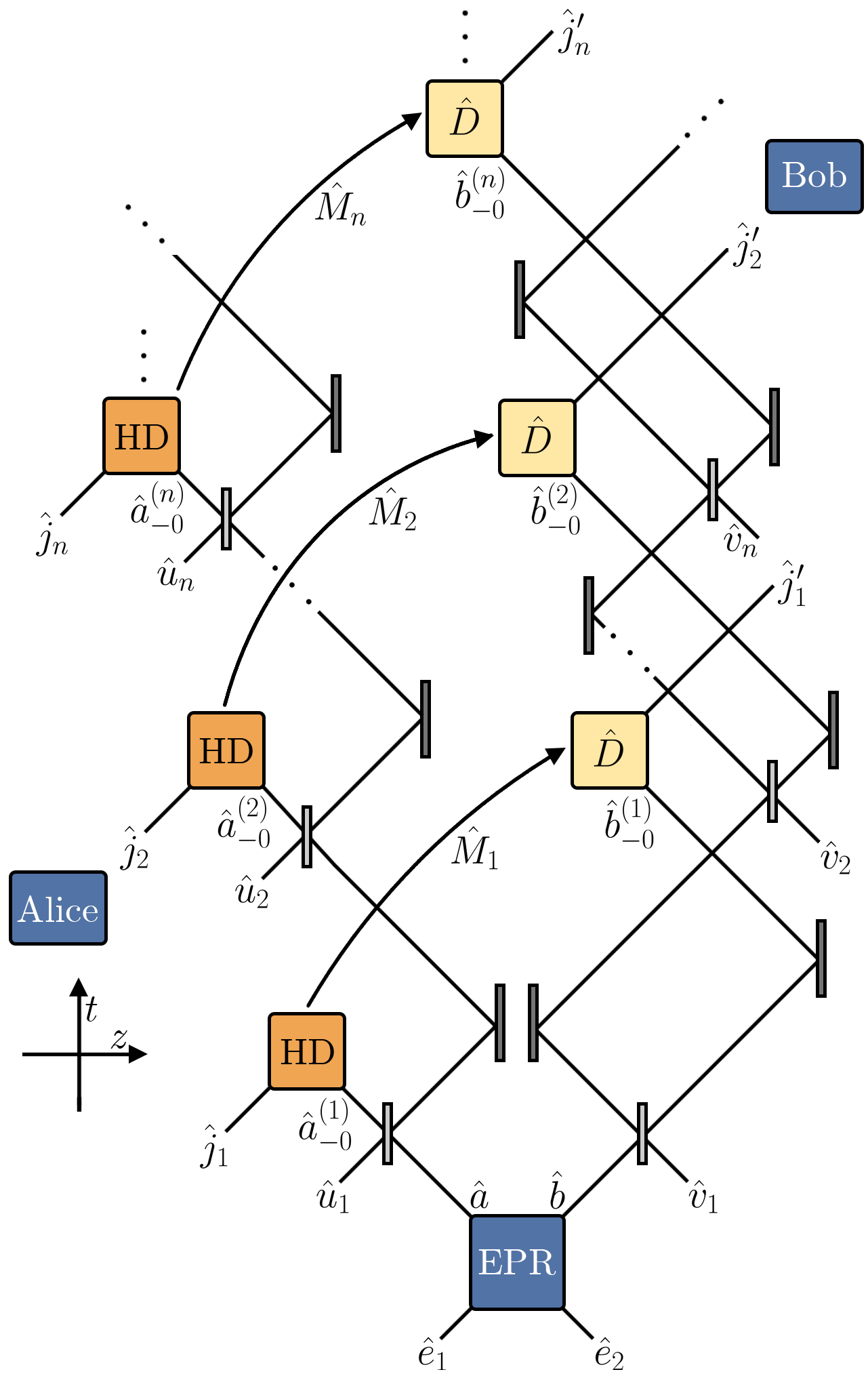}
    \caption{Circuit diagram for the $N$-mode temporal mode telefilter with no time-delay. }
    \label{fig:nmodenodelay}
\end{figure}
Our entanglement is split up in the same way as the time-delayed telefilter case. As before, let us specialise to $\phi_i = - \pi/2$ for the beamsplitter phases, for simplicity. We can construct the individual measurement results from the quadrature operators obtained at each homodyne measurement, as follows:
\begin{align}
    \hat{M}_1 &= |\beta|( \hat{X}_{j_{10}'} + i \hat{P}_{a_{-0}^{(1)\prime} }) = \sqrt{2}|\beta| ( \hat{j}_{10} + \sqrt{1 - \alpha_1} \hat{a}_0^\dd + \sqrt{\alpha_1} \hat{v}_{10} )  \vt \\
    \hat{M}_2 &= |\beta| (\hat{X}_{j_{20}'} + i \hat{P}_{a_{-0}^{(2)\prime}})  = \sqrt{2} |\beta| ( \hat{j}_{20} + \sqrt{\alpha_1( - \alpha_2) } \hat{a}_0^\dd - \sqrt{(1 - \alpha_1 ) (1 - \alpha_2)} \hat{v}_{10}^\dd + \sqrt{\alpha_2} \hat{v}_{20}^\dd )  \vt \\
    & \:\:\:\vdots \non \\
    \hat{M}_n &= |\beta| ( \hat{X}_{j_{(n)0}'} + i \hat{P}_{a_{-0}^{(n)\prime}}) = \sqrt{2} |\beta| ( \hat{j}_{n0} + \sqrt{\alpha_1 \alpha_2 \hdots \alpha_{n-1} ( 1- \alpha_n)} \hat{a}_0^\dd - \sqrt{(1 - \alpha_1) \alpha_2 \hdots \alpha_{n-1} ( 1- \alpha_n) } \hat{v}_{10}^\dd \non \vt \\
    & - \sqrt{(1 - \alpha_2) \alpha_3 \hdots \alpha_{n-1} (1- \alpha_n)} \hat{v}_{20}^\dd - \hdots - \sqrt{(1 - \alpha_{n-1} ) ( 1- \alpha_n)} \hat{v}_{(n-1)0}^\dd + \sqrt{\alpha_n} \hat{v}_{(n)0}^\dd ) \vt  \\
    & \:\:\:\vdots \non \\ 
    \hat{M}_{N} &= |\beta| (\hat{X}_{j_{N0}'} + i \hat{P}_{a_{+0}^{(N-1)\prime}})  = \sqrt{2}|\beta| ( \hat{j}_{(N)0} + \sqrt{\alpha_1 \hdots \alpha_{N-1}} \hat{a}_0^\dd - \sqrt{(1 - \alpha_1) \alpha_2 \hdots \alpha_{N-1}} \hat{v}_{10}^\dd - \sqrt{(1 - \alpha_2) \alpha_3 \hdots \alpha_{N-1}} \hat{v}_{20}^\dd \non \vt  \\
    & - \sqrt{(1 -\alpha_{N-2} )\alpha_{N-1} } \hat{v}_{(N-2)0}^\dd - \sqrt{1 - \alpha_{N-1}} \hat{v}_{(N-1)0}^\dd \vt 
\end{align}
We use these measurement results to displace Bob's half of the entanglement resource modes. We choose the effective gain of the displacement to be $\zeta_i = - 1/(\sqrt{2}
\beta|)$ which yields an identity channel between Alice and Bob for each of the temporal modes: 
\begin{align}
    \hat{j}_{10}' &= - \hat{j}_{10} + \sqrt{1 - \alpha_1} (\hat{b}_0  - \hat{a}_0^\dd ) + \sqrt{\alpha_1} ( \hat{u}_{10} - \hat{v}_{10}^\dd )  \vt \\
    \hat{j}_{20}' &= - \hat{j}_{20} + \sqrt{\alpha_1 ( 1- \alpha_2)} (\hat{b}_0 - \hat{a}_0^\dd ) - \sqrt{(1- \alpha_1) ( 1- \alpha_2)} ( \hat{u}_{10} - \hat{v}_{10}^\dd ) + \sqrt{\alpha_2} ( \hat{u}_{20} - \hat{v}_{20}^\dd ) \vt  \\
    & \:\:\: \vdots \non \\
    \hat{j}_{(n)0}' &= - \hat{j}_{(n)0} + \sqrt{\alpha_1 \alpha_2 \hdots (1 -\alpha_n)} ( \hat{b}_0 - \hat{a}_0^\dd ) - \sqrt{(1 - \alpha_1 ) \alpha_2 \hdots \alpha_{n-1} (1-\alpha_n) } ( \hat{u}_{10} - \hat{v}_{10}^\dd ) \non \vt \\
    & - \sqrt{(1 - \alpha_2 )\alpha_3 \hdots \alpha_{n-1} ( 1- \alpha_n) }( \hat{u}_{20} - \hat{v}_{20}^\dd ) - \hdots - \sqrt{(1-\alpha_{n-1}) (1 - \alpha_n) } ( \hat{u}_{(n-1)0} - \hat{v}_{(n-1)0}^\dd )+ \sqrt{\alpha_n} ( \hat{u}_{(n)0} - \hat{v}_{(n)0}^\dd ) \vt \\
    & \:\:\: \vdots \non \\
    \hat{j}_{(N)0}' &= - \hat{j}_{(N)0 } - \sqrt{(1 - \alpha_1 ) \alpha_2 \hdots \alpha_{N-1} } (\hat{u}_{10} - \hat{v}_{10}^\dd ) - \sqrt{(1 - \alpha_2 ) \alpha_3 \hdots \alpha_{N-1}} ( \hat{u}_{20} - \hat{v}_{20}^\dd ) \non \vt \\
    & - \hdots - \sqrt{(1 - \alpha_{N-2} \alpha_{N-1}} ( \hat{u}_{(N-2)0} - \hat{v}_{(N-2)0}^\dd ) - \sqrt{1 - \alpha_{N-1}} (\hat{u}_{(N-1)0} - v_{(N-1)0}^\dd )  \vt 
\end{align}
Performing a similar recombination of the output modes on passive beamsplitters (which are weighted in with respect to the initial distribution of the entanglement resource modes across different temporal modes), we finally obtain
\begin{align}
    \hat{o}_{10}^{(l)} &= - \sqrt{\alpha_{N-1}} \hat{j}_{(N-1)0} + \sqrt{1 - \alpha_{N-1}} \hat{j}_{(N)0 } + \underbrace{( \hat{u}_{(N-1) 0 } - v_{(N-1)0}^\dd ) }_\text{noise} \\
    \hat{o}_{20}^{(l)} &= - \sqrt{\alpha_{N-2}} \hat{j}_{(N-2) 0 } + \sqrt{(1 - \alpha_{N-2})( 1- \alpha_{N-1})} \hat{j}_{(N-1) 0 } + \sqrt{(1 - \alpha_{N-2}) \alpha_{N-1} } \hat{j}_{(N)0 } + \underbrace{( \hat{u}_{(N-2)0} - \hat{v}_{(N-2)0}^\dd )}_\text{noise}  \\
    & \:\:\: \vdots \non \\
    \hat{o}_{(n)0}^{(l)} &= - \sqrt{\alpha_{N-n} } \hat{j}_{(N-n)0} + \sqrt{(1 - \alpha_{N-n})( 1- \alpha_{N-(n-1)})} \hat{j}_{(N-(n-1))0} \non \\
    & + \hdots + \sqrt{(1 - \alpha_{N-n}) \alpha_{N-(n-1) }\alpha_{N-(n-2)} \hdots \alpha_{N-3} ( 1- \alpha_{N-2})} \hat{j}_{(N-2)0} + \sqrt{(1 - \alpha_{N-n}) \alpha_{N-(n-1)} \alpha_{N-(n-2)} \hdots \alpha_{N-2} ( 1- \alpha_{N-1} )  } \hat{j}_{(N-1)0} \non \\
    & + \sqrt{(1 - \alpha_{N-n} )  \alpha_{N-(n-1) } \alpha_{N-(n-2)} \hdots \alpha_{N-2} ( 1 - \alpha_{N-1}) }\hat{j}_{(N-1)0} \non \\
    & + \sqrt{(1 - \alpha_{N-n} ) \alpha_{N-(n-1)} \alpha_{N-(n-2)} \hdots \alpha_{N-1}} \hat{j}_{(N)0} + ( \underbrace{\hat{u}_{(N-n)0} - \hat{v}_{(N-n)0}^\dd )}_\text{noise} \non \\
    \hat{o}_{(N-1)0}^{(r)} &= - \sqrt{1- \alpha} \hat{j}_{10} - \sqrt{\alpha_1 (1- \alpha_2 ) }\hat{j}_{20} - \sqrt{\alpha_1 \alpha_2 (1 - \alpha_3)} \hat{j}_{30} - \hdots - \sqrt{\alpha_1 \alpha_2 \hdots \alpha_{N-2} ( 1- \alpha_{N-1} ) } \hat{j}_{(N-1) 0 } \non \\
    & - \sqrt{\alpha_1 \alpha_2 \hdots \alpha_{N-2} \alpha_{N-1}} \hat{j}_{(N)0 } + \underbrace{( \hat{b}_0 - \hat{a}_0^\dd ) }_\text{entanglement}
\end{align}
Here, the discriminated mode is contained in $\hat{o}_{(N-1)0}^{(r)}$. For this choice of beamsplitter settings, this is a superposition of the $\hat{j}_{i0}$ temporal modes with equal phases between them, along with the entanglement resource. Meanwhile, the other $N$ output modes contain orthogonal superposition modes with noise. 
\end{widetext}

\bibliography{main}

\end{document}